\documentclass[]{aa}
\usepackage[varg]{txfonts}
\usepackage{graphicx}
\usepackage{natbib}
\usepackage{color}

\begin{document}

\title{The age structure of stellar populations in the solar vicinity}
\subtitle{Clues of a two-phase formation history of the Milky Way disk}

\author{Misha Haywood\inst{1}
  \and Paola Di Matteo\inst{1}
     \and Matthew D. Lehnert\inst{2}
  \and David Katz\inst{1}
  \and Ana G\'omez\inst{1}
}
\authorrunning{Haywood et al.}

\offprints{M. Haywood, \email{Misha.Haywood@obspm.fr}}

\institute{GEPI, Observatoire de Paris, CNRS, Universit\'e
  Paris Diderot, 5 place Jules Janssen, 92190 Meudon, France\\
\and 
Institut d'Astrophysique de Paris, Universit\'e Pierre et Marie Curie/CNRS,98 bis Bd Arago, F-75014 Paris, France
}

\date{Received / Accepted }

\abstract{We analyze a sample of solar neighborhood stars which have
high-quality abundance determinations and show that there are two distinct
regimes of [$\alpha$/Fe] versus age which we identify as the epochs of the
thick and thin disk formation. A tight correlation between metallicity and
[$\alpha$/Fe] versus age is clearly identifiable for thick disk stars,
implying that this population formed from a well mixed interstellar
medium, probably initially in starburst and then more quiescently,
over a time scale of 4-5 Gyr. Thick disk stars have vertical velocity
dispersions which correlate with age, with the youngest objects of this
population having small scale heights similar to that of thin disk
stars.  A natural consequence of these two results is that a vertical
metallicity gradient is expected in this population. We suggest that
the youngest thick disk set the initial conditions from which the {\it
inner} thin disk started to form about 8 Gyr ago, at [Fe/H] in the range of
(-0.1,+0.1) dex and [$\alpha$/Fe]$\sim$0.1 dex.  This also provides
an explanation of the apparent coincidence between the existence of a
step in metallicity at 7-10~kpc in the thin disk and the confinement
of the thick disk within R$<$10~kpc. We suggest that the {\it outer}
thin disk developed outside the influence of the thick disk, giving
rise to a separate structure, but also that the high alpha-enrichment
of those regions may originate from a primordial pollution of
the outer regions by the gas expelled from the forming thick disk.
Metal-poor thin disk stars ([Fe/H]$<$-0.4 dex) in the solar vicinity,
whose properties are best explained by them originating in the outer
disk, are shown to be as old as the youngest thick disk (9-10~Gyr). This
implies that the outer thin disk started to form while the thick disk
was still forming stars in the inner parts of the Galaxy.  Hence, while
the overall inner (thick+thin) disks is comprised of two structures with
different scale lengths and whose combination may give the impression
of an inside-out formation process, the local thin disk itself probably
formed its first stars in its outskirts.  Moreover, we point out that,
given the tight age-metallicity and age-[$\alpha$/Fe] relations that
exist in the thick disk, an inside-out process would give rise to a radial
gradient in metallicity and $\alpha$-elements in this population which is
not observed. Finally, we argue that our results leave little room for
radial migration (in the sense of churning) either to have contaminated
the solar vicinity, or, on a larger scale, to have redistributed stars
in significant proportion across the solar annulus.}


\maketitle

\section{Introduction} 

A stellar population is commonly viewed as an ensemble of stars that
share some observable properties \citep[][]{baa44, mcc59}. A subtler
definition would allow us to link some observable quantities to properties
that we think uniquely define the conditions of the formation of the
population. Hopefully, such an approach would aide us in finding
out how the formation of individual stellar populations relates
to the evolution of the galaxies generally.  In practice, however,
such link is difficult to tease out because there is no simple way to
define unambiguously a single population of stars.  An example of this
failure is epitomized by the recent discussion about the existence of
a thick disk as a separate population in its own right
\citep[e.g.][]{bov12a, bov12b}.  Structural parameters may give us a hint of the existence
of such a population in the Milky Way \citep{gil83} as well
as in external galaxies \citep{bur79, dal02, poh04, com11}, but no clue as to its
origin or to its link to the other populations. Kinematics are also often
used to segregate thin and thick disk stars \citep[e.g.][]{sou03, ben03}, but 
have currently offered few insights beyond
simple classification, and may even represent a loss of information
if it is used to group all the kinematic data.  Finally, chemical
characteristics have been proposed as a way to parametrize the properties
of various stellar populations in the Milky Way \citep[scale height, scale length,
kinematics, etc. see ][]{bov12b}, or for setting limits on the
characteristics of the thin and thick disks \citep[see][]{nav11},
but defined in an arbitrary way, and with no direct evidence that these
definitions correspond to the properties of the interstellar medium
(ISM) at a particular epoch.  Up to now, no clear picture has emerged
as to where to put boundaries or even if boundaries exist between the
thick and thin disk population \citep{nor87,bov12b}.

In the present study, we analyze the properties of a sample of FGK stars
within the solar vicinity. We show that the formation of stars within the
thick disk is clearly identifiable as a phase of steady and monotonic
variation in the chemical properties over a period of about 4-5 Gyr.
We suggest that the processes that occurred during the formation of
the thick disk set the initial conditions for the formation of the
inner thin disk, while the outer disk may have built up independently
but certainly over a longer time scale.

In the following section, we describe the sample used for this
analysis. In \S 3, we give a description of the ``age structure'' of
the two disk populations through an inspection of HR diagrams and the
determination of individual ages.  We outline the implications of our
results in \S 4, propose a scenario for the formation of the Milky Way
disk in \S 5, and discuss some critical aspects of disk formation in
the light of our findings. Finally, we summarize our results in \S 6.

\section{Sample characteristics and derived ages} 

\begin{figure}
\includegraphics[width=9cm]{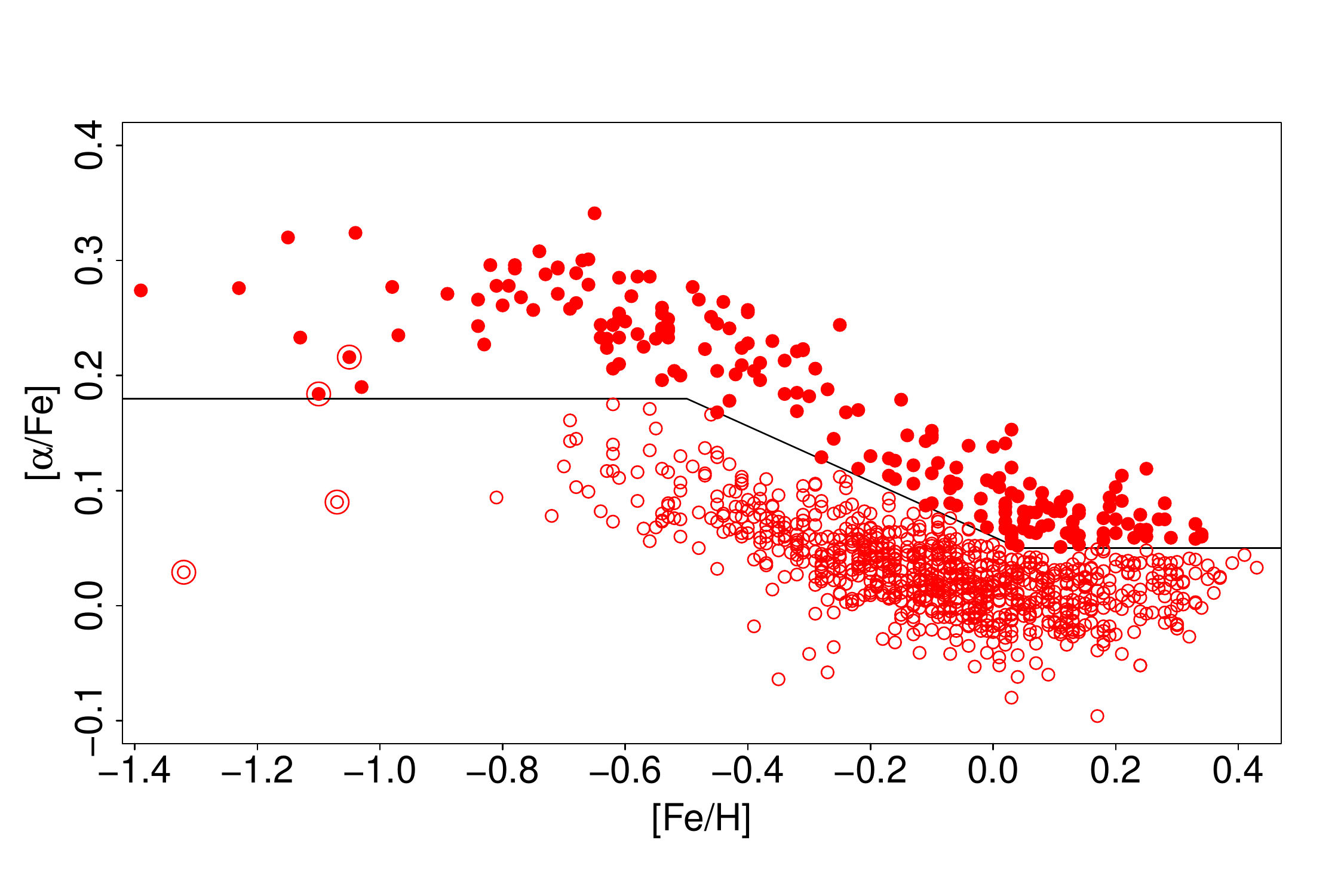}
\caption{[$\alpha$/Fe] $vs$ [Fe/H] for stars in the sample of Adibekyan
et al. (2012).  $\alpha$ is defined as the mean of the abundances
of magnesium, silicon and titanium. The thick and thin disk sequences
are distinguished by whether they lie above or below the three segments of
the solid line respectively.
The two doubled circles at [Fe/H]$<$-1.0 dex and [$\alpha$/Fe] $<$ 0.1 dex are 
HIP 74234 and HIP 100568, while the two dot+circle symbols represent 
HIP 54641 and HIP 57360 (see text).
}
\label{alphafeh}
\end{figure}

The sample used in this study was originally designed for the detection
of planets through radial velocity measurements (see \citet{adi12} for a description). It comprises 1111 FGK stars with spectra
at a resolution of R=110000 and excellent S/N ratio for most objects
(the authors report a S/N better than 200 for 55\% of the sample).
The sample is composed of various subsamples defined on the basis of
spectral type range (FGK), limiting distance \citep[see][]{udr00, loc10}, low rotational velocity, and low atmospheric activity. 97
stars in the sample were selected to have photometric metallicities
between -0.5 and -1.5 and b-y$>$0.33. To our knowledge, no other cut in
colour or kinematics were made in the selection of the stars.

Atmospheric parameters (T$_{eff}$, [Fe/H], [$\alpha$/Fe]) are adopted from
\citet{adi12}, while absolute magnitudes were calculated from
the parallaxes presented in \citet{van07}\footnote{No reddening correction have been
applied, 95\% of the sample being at distances nearer than 75pc}.  Note that we have
defined the $\alpha$ element abundance as the mean of magnesium, silicon
and titanium, while excluding calcium, as recommended in \citet{adi12}.
We adopt galactic space UVW velocities \citep[corrected for the solar
motion using the values from][]{sch10} from the study of \citet{adi12}.
Further, we determined the orbital parameters from orbits
calculated in a Galactic axisymmetric potential \citep{all91}.
We show the sample of stars in the [$\alpha$/Fe] $vs$
[Fe/H] plane in Fig. \ref{alphafeh}. The upper sequence (above the
line) is usually classified as containing thick disk stars, and stars
in the lower sequence are classified as thin disk. However, in such a
scheme, the classification of the most metal-rich stars on
the thick disk sequence is not clear. Note that the two deficient stars
([Fe/H]$<$-1.0 dex) with low alpha content ([$\alpha$/Fe]$<$+0.1 dex)
are stars which are counter rotating with respect to the disk. 
The two other deficient stars identified on Fig. \ref{alphafeh} and which have 
[$\alpha$/Fe]$\approx$+0.2 dex, HIP 54641 and HIP 57360,
have a strong U component and are significantly 
younger (see Fig. \ref{agemet}) than typical stars of similar metallicity.
All four stars are possible members of the so-called accreted halo \citep[see][]{nis10}\protect\footnote{We warmly thank Poul Nissen for suggesting to us to investigate more carefully at the properties of these stars.}.

We derived individual ages using the bayesian method of \citet{jor05}, adopting the Yonsei-Yale (Y$^2$) set of isochrones \citep[version 2,][]{dem04}. Atmospheric parameters were taken from the study of
\citet{adi12}, including [$\alpha$/Fe] values, as defined in the
previous paragraph.  

\subsection{Uncertainties on ages due to atmospheric parameters}

Formal errors in the individual atmospheric parameters \citep{adi12} are less than 70~K in the effective temperature ($<$25~K for
solar-type stars), $<$0.05~dex for [Fe/H] ($<$~0.02 dex for solar-like
stars) and $<$0.1~dex in [$\alpha$/Fe] ($<$0.03~dex for solar-type
stars).  Adopting such small uncertainties in the atmospheric parameters
would imply unrealistically robust ages, so we adopted conservative
uncertainties of 50K in temperature, 0.1 dex in metallicity and 0.1 mag
in absolute magnitude. These uncertainties in the atmospheric parameters
translate into uncertainties of about 0.8 Gyr for the younger stars
(age $<$ 5~Gyr) and 1.5~Gyr for the oldest (age $>$ 9~Gyr).  While random
errors can be important for individual objects, systematic uncertainties
that can affect these parameters (or stellar models) may result in very
significant biases in the derived ages and perhaps lead to spurious trends
\citep{hay06}. \citet{sou11} compared stars that are in common
with this sample and various other samples \citep{edv93,ben03,val05}, showing that there
is no systematic difference in these analyses for [Fe/H] or T$_{eff}$
for stars with T$_{eff}<$6000K, but found a significant offset of about
100K above this T$_{eff}$ limit for stars in common with the \citet{val05}
sample.

\begin{figure}
\includegraphics[width=9cm]{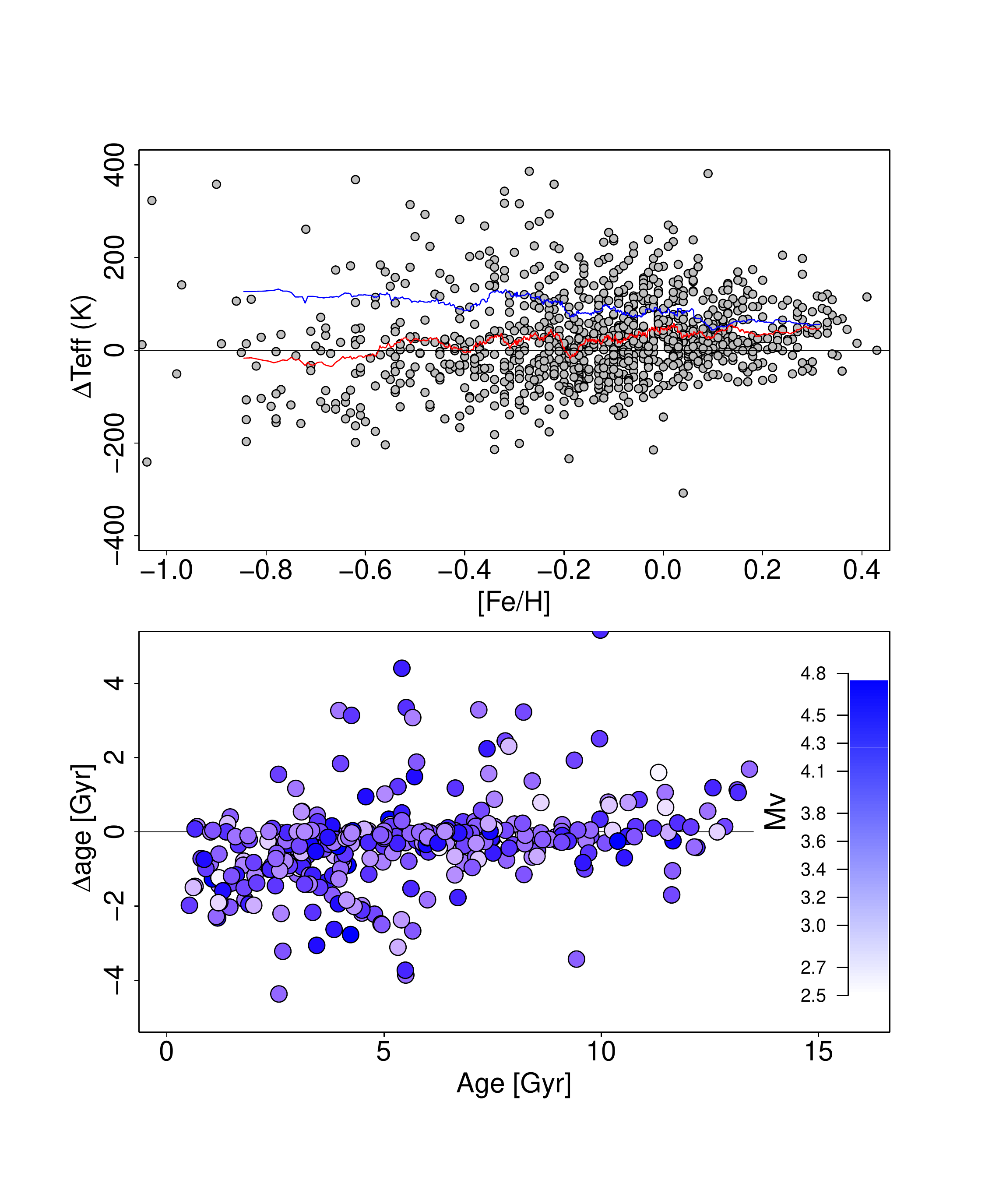}
\caption{Upper plot: Difference between effective temperatures from Adibekyan et al. (2012) and those
derived from the V-K indices and calibration from Casagrande et al. (2010) for the whole sample. 
Lower curve is the mean of the difference calculated over 75 points, the upper curve is the dispersion.
Lower plot: Difference in age resulting from the two scales, as a function of age (based on Adibekyan 
et al. (2012) spectroscopic scale) for stars with `good' ages (see text).
}
\label{agetest}
\end{figure}

\begin{figure*}
\includegraphics[width=19cm]{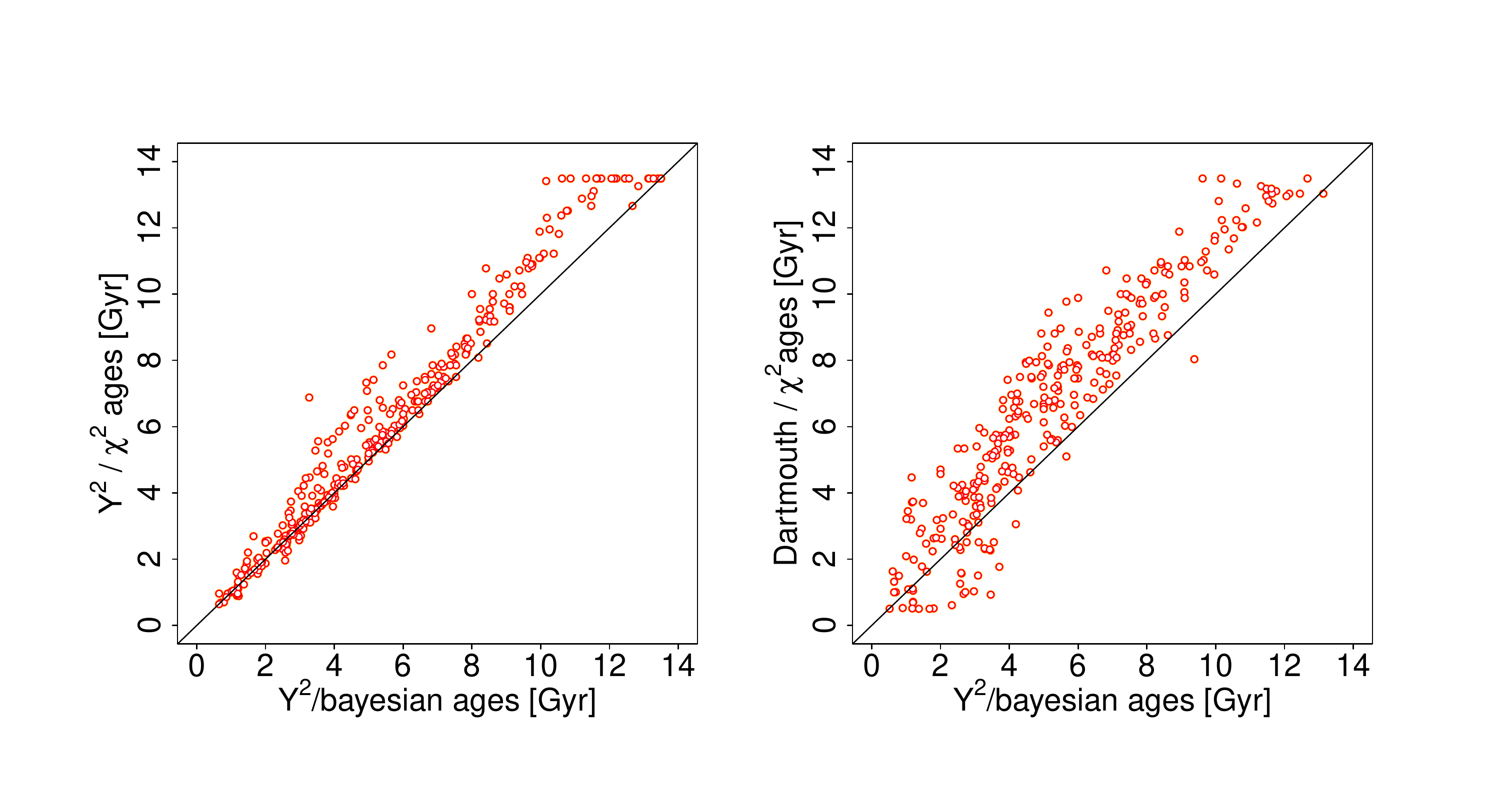}
\caption{Left: $\chi^2$ $vs$ bayesians ages, using the same set of isochrones. Right: $\chi^2$ (Dartmouth isochrones) $vs$ bayesians ages (Y$^2$ isochrones). 
}
\label{isoaccuracy}
\end{figure*}

Since systematic uncertainties that affect the temperature scale can produce
the most significant effects on the age determinations \citep{hay06}, we 
compared the spectroscopic temperatures \citep{adi12} with those derived from the V-K color index plus a
state-of-the-art calibration \citep{cas10}.  The comparison
(Fig. \ref{agetest}) shows that the mean of the
differences (running average calculated over 75 stars) between the
spectroscopically-derived and the V-K color temperature is always lower
than 56~K. These differences are in the sense that
spectroscopic temperatures are higher. The agreement is remarkable
given that the two scales are entirely independent and suggests that
the systematic uncertainties in the effective temperatures are actually
quite low.  

In order to illustrate the effect of these minor offsets and random
uncertainties in effective temperatures, we derived ages using
temperatures from the V-K color. Fig. \ref{agetest} (bottom plot) shows the age 
difference (age on Adibekyan T$_{eff}$ - age from V-K T$_{eff}$ scale) between the two determinations as a function of age 
(on Adibekyan et al. T$_{eff}$ scale).
70\% of the stars have an age difference lower than 1 Gyr, and below 0.5 Gyr for 50\% of the cases. 
For stars with ages greater than 5 Gyr, 50\% have an age difference lower than 0.5 Gyr.

\subsection{Uncertainties on ages due to the method and to the stellar physics}

As a final test to the robustness of our ages, we did two additional checks:
we first derived ages applying a simple $\chi^2$ minimization between the position of the star and 
the nearest isochrone (at the metallicity and [$\alpha$/Fe] of the star) in the (effective temperature, absolute 
magnitude) plane, following the prescription from \citet{ng98} and using Y$^2$ isochrones.
The result is shown in Fig. \ref{isoaccuracy} (left), with $\chi^2$-ages as a function of bayesian ages. 
There is a trend in the sense that $\chi^2$ minimization gives
greater ages than the bayesian method for the older stars, reaching a difference of about 1.5 Gyr 
at ages $>$13 Gyr. We think this is due to the fact that the  $\chi^2$ minimization
favors older isochrones, which are more closely spaced for a fixed age step. 
While the relative merit of each method is beyond the scope of the present analysis, we note that 
apart from this systematic shift, small random uncertainty (of the order of 0.7 Gyr) also arises 
due to difference in the way the
'best' age is defined in each method. In particular, the stretch of stars which deviate 
gradually from the diagonal starting at 4 Gyr upward, is due to the stars in the hook region, 
and whose bayesian age shows a doubled-peaked probability distribution function. 
 
The second check is made deriving new ages using the $\chi^2$ minimization and the set of isochrones 
from Dotter et al. (2007; Dartmouth isochrones). The two age scales are compared in 
Fig. \ref{isoaccuracy} (right panel).
The plot illustrates that uncertainties in stellar physics add both a systematic and 'random' 
error to the ages: in addition to the shift due to the method, Dartmouth isochrones 
give systematically older ages by about 1 Gyr. A 'random' uncertainty also adds up, which 
amounts to almost 1 Gyr. 
Therefore, we expect the absolute age scale used in this study to be  uncertain by about
1 to 1.5 Gyr, while relative ages have uncertainties of about 1 Gyr, due to  
stellar physics.

From the initial sample of 1111 stars, about half could be age dated. After taking into account stars with a well defined probability 
distribution function together
with absolute magnitude Mv$<$4.75, the final subsample with meaningful ages has 363
stars, a rather severe pruning, but which ensures robust age estimates.

\section{Results} \label{res}

For convenience, hereafter we 
adopt the terms \emph{thin and thick disk
sequences} to designate stars that are below or above the line in
Fig.~\ref{alphafeh}.  This definition will be then compared to that  of \emph{thin and thick disk populations},
introduced in \S~\ref{individual}.

\subsection{General trends in age: H-R diagrams}

\begin{figure*}
\centering
\includegraphics[width=17cm]{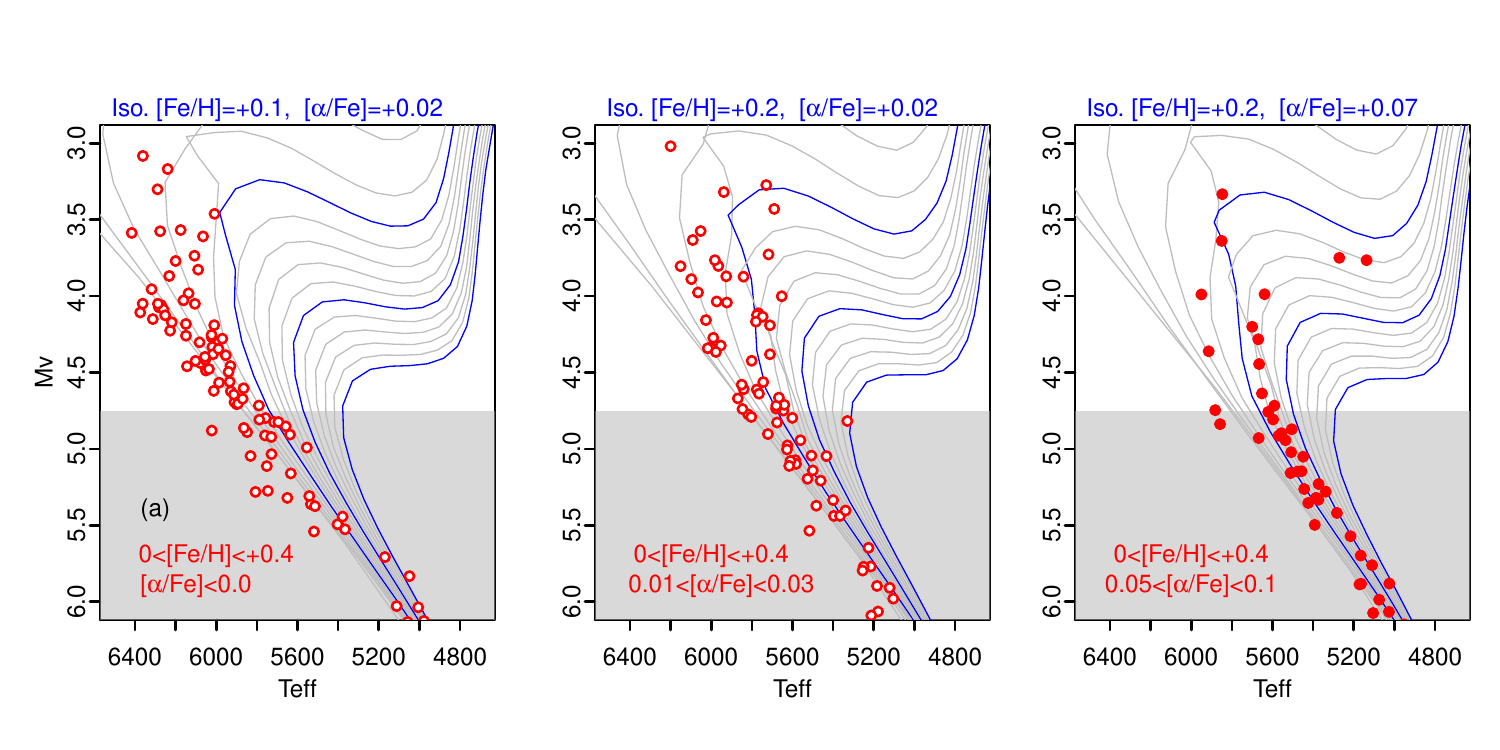} 
\includegraphics[width=17cm]{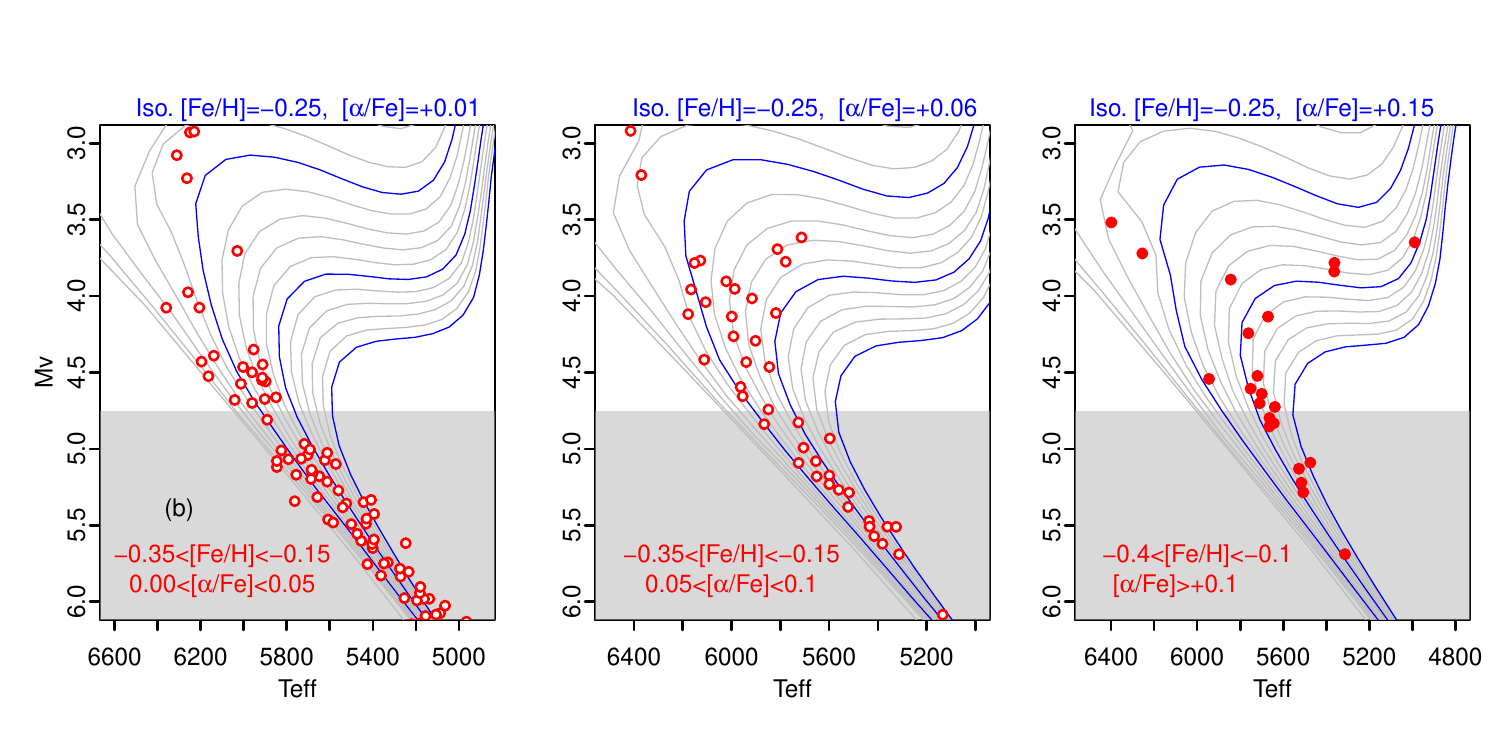} 
\includegraphics[width=17cm]{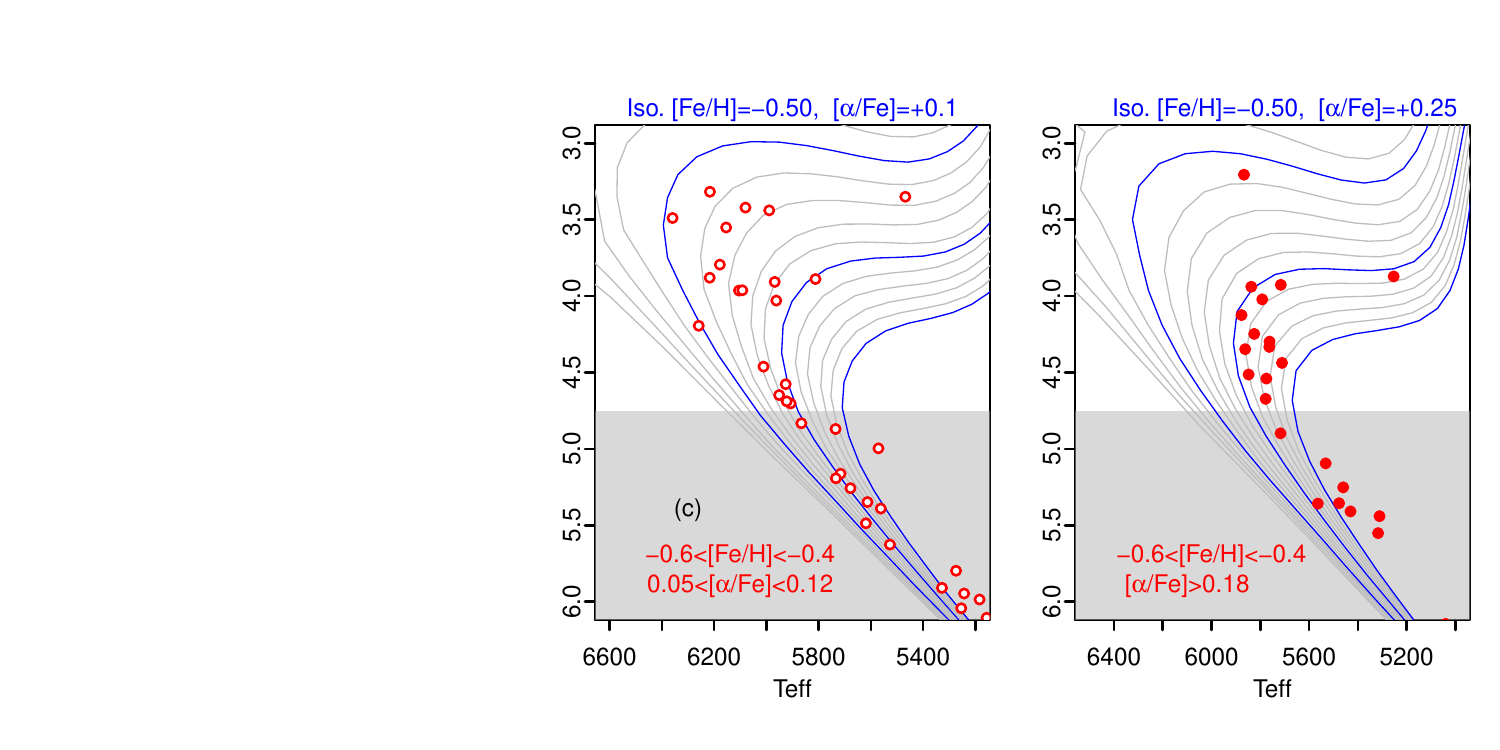} 
\caption{H-R diagrams for stars in various metallicity ranges
and [$\alpha$/Fe] content (given
explicitly in the legend at the bottom left corner of each plot).
The lines in each plot are isochrones of various ages (Demarque et
al. 2004). Isochrones shown as solid dark blue lines have ages of 5,
10 and 15 Gyr (from left to right). The open and solid red circles are
the same as defined in Fig. 1. Stars that have been dated are above the grey area.}
\label{dhrtndisk}
\end{figure*}

\begin{figure*}
\includegraphics{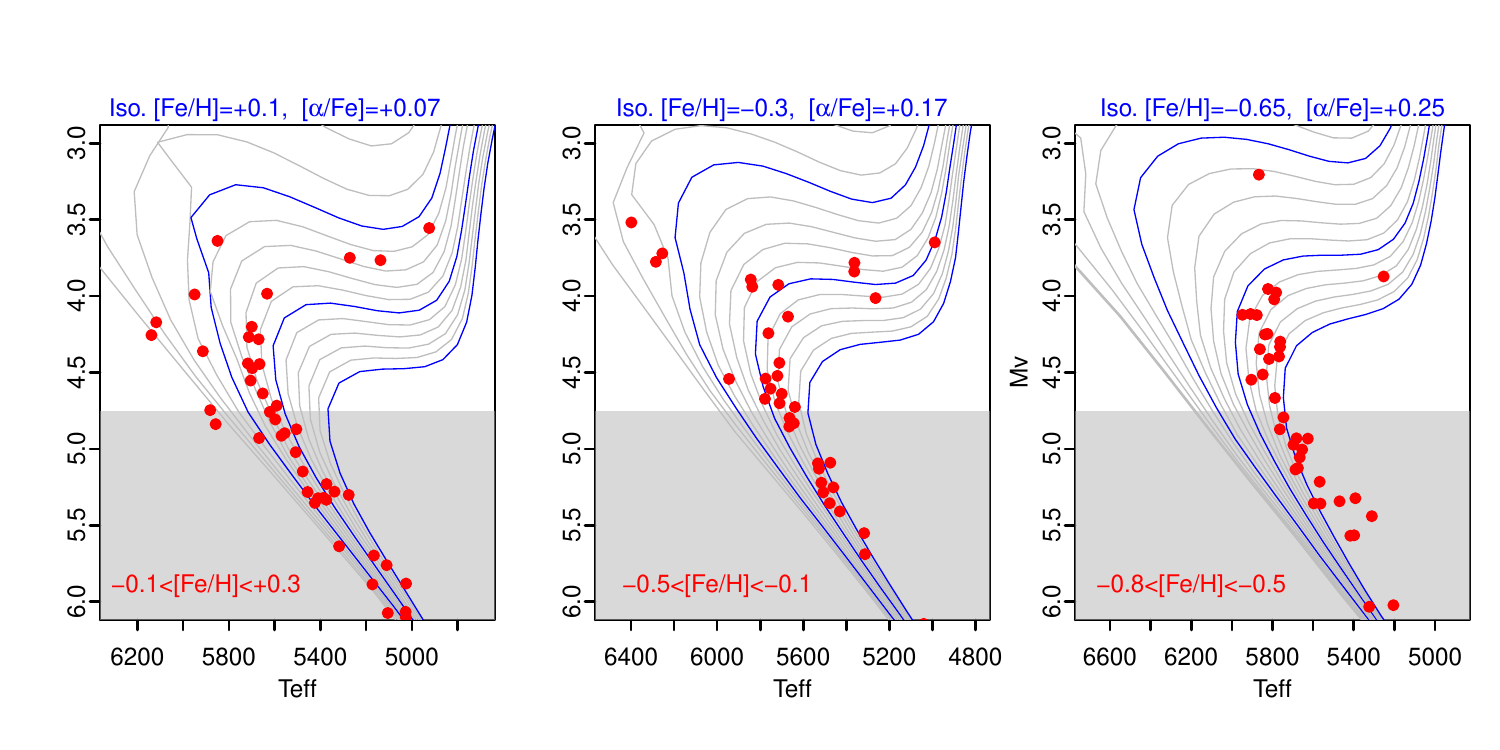} 
\includegraphics{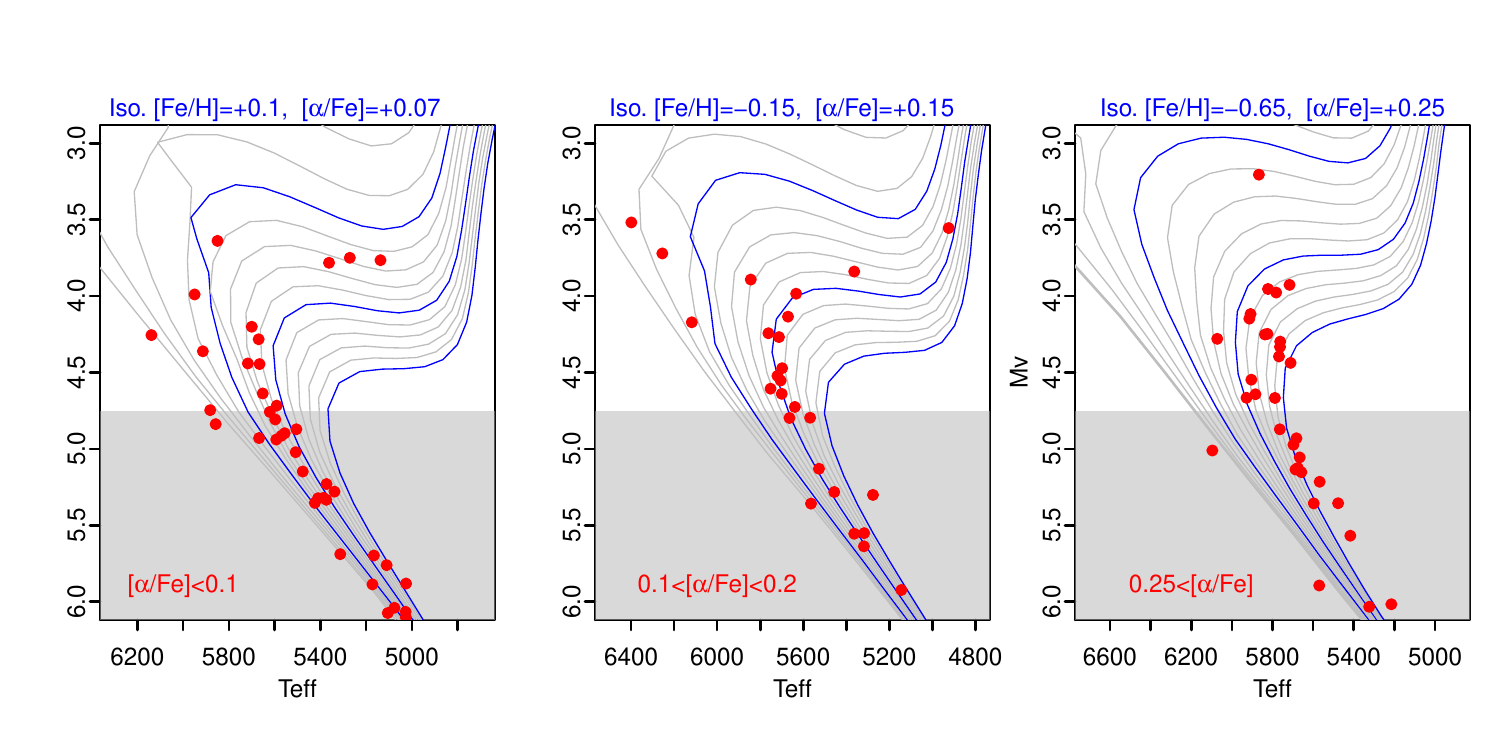} 
\caption{H-R diagrams for stars selected in the thick disk sequence for
various ranges in metallicity.  The exact range
in [Fe/H] is indicated in the legend in the bottom left
corner of each plot. The isochrones are the same as in Fig.~\ref{dhrtndisk}.
Stars that have been dated are above the grey area.}
\label{dhrtckdisk}
\end{figure*}

Before presenting the relations between age, chemistry and kinematics, in this section we investigate
the trends in the variation
of [$\alpha$/Fe] with age by simple inspection of H-R diagrams (Fig.~\ref{dhrtndisk}). Such
a simple approach has the advantage of not relying on methods used to
determine individual ages and their concomitant systematic uncertainties.
At [$\alpha$/Fe]$<$0.0~dex, most objects are younger than 3~Gyr, while
they lie between 3 and 6~Gyr for 0.01$<$[$\alpha$/Fe]$<$0.03~dex
and 5-7 Gyr for [$\alpha$/Fe]$>$0.06~dex (Fig.~\ref{dhrtndisk}).
The middle row of Fig.~\ref{dhrtndisk} shows a similar sequence for
stars in the metallicity range $\sim$ $-$0.4$<$[Fe/H]$<$$-$0.1~dex.  Note that
there is no significant number of stars with such metallicities and
[$\alpha$/Fe]$<$0.0~dex.  At  0.00$<$[$\alpha$/Fe]$<$0.05, stars are
grouped around 4-5~Gyr, as in the upper row, middle plot. In the  range
0.05$<$[$\alpha$/Fe]$<$0.1, stars are bracketed between the isochrones
of 5 and 10 Gyr as it is the case for stars with the same range of alpha
abundance and higher metallicities (upper row, Fig.~\ref{dhrtndisk}).
Finally, in the last plot of the middle row, only a few stars with
[$\alpha$/Fe]$>$0.1 are seen which lie around the 10 Gyr isochrone. The
bottom row shows stars at $-$0.6$<$[Fe/H]$<$$-$0.4~dex, which mostly
lie between the 5 and 8 Gyr isochrones for 0.05$<$[$\alpha$/Fe]$<$0.12,
and beyond the 10 Gyr isochrone for  $[\alpha/Fe]>$0.18 dex. Clearly,
there is a good qualitative relationship between $\alpha$ enhancement
and age over the various metallicity intervals.

We now repeat that same exercise but selecting the objects that lie
along the sequence of the thick disk stars (i.e., above the separating
line in Fig. \ref{alphafeh}).  The upper row in Fig. \ref{dhrtckdisk}
shows first the position of the stars in the H-R diagram as a function of
metallicity, while the bottom row shows the positions of stars in the H-R
diagram for three different alpha abundance intervals.  Interestingly,
both variations in metallicity and in [$\alpha$/Fe] abundance show a
variation with age in the sense that the most metal-rich stars have ages
around 7-8~Gyr, shifting to about 10 Gyr around [$\alpha$/Fe]=0.15~dex
and [Fe/H]=-0.15~dex, and 11-13~Gyr at [$\alpha$/Fe]$>$0.25~dex.
These results indicate a real correlation between age and both
[$\alpha$/Fe] and metallicity for stars in the thick disk sequence.
Whether all these stars belong to a ``thick disk population'' is the
subject we will discuss subsequently.

\subsection{The age-[$\alpha$/Fe] relation: towards defining the thin
and thick disk populations from ages}\label{individual}

\begin{figure}
\centering
\includegraphics[width=9.5cm]{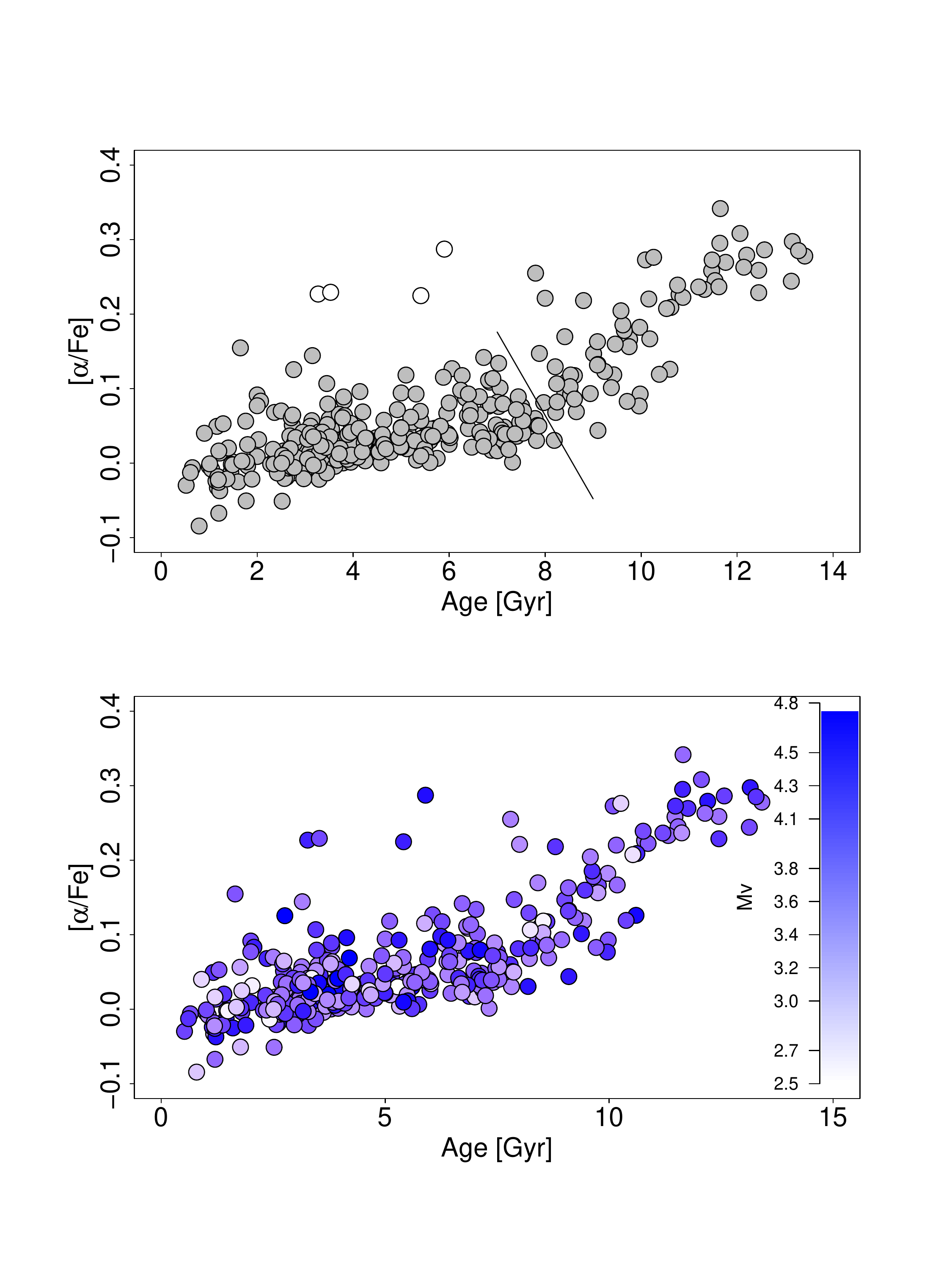}
\caption{{\it Top panel:} The general [$\alpha$/Fe] $vs$ age distribution
for stars in the sample, for all stars having an absolute magnitude
M$_v$ $<$ 4.75.  Open circles indicate stars which
have alpha enhancements consistent with the thick disk but have ages
indicative of the thin disk and also lie on the thick disk sequence in the
[$\alpha$/Fe]--[Fe/H] plane.  The line divides the thick and thin disk populations.
{\it Bottom panel:} Same stars, 
the color coding the absolute magnitude of stars as described by the vertical
scale.
}
\label{agealpha2}

\end{figure}

A logical place to start when attempting to define stellar populations is
by investigating relationships with their ages and other characteristics
that may act like chronometers, such as the relative alpha enhancement,
[$\alpha$/Fe].  The upper panel of Fig.~\ref{agealpha2} shows a
correlation between [$\alpha$/Fe] and age such that the [$\alpha$/Fe]
enhancement of stars decreases with the stellar age.  Interestingly,
the relation suggests that the rate at which the alpha enhancement
declines, $\Delta[\alpha/Fe]/\Delta t$, has a break at an age of
$\sim$8~Gyr. $\Delta[\alpha/Fe]/\Delta t$ is a factor of 5 higher for
stars older than $\sim$8~Gyr compared to stars which are younger.
There are few outliers in the age-[$\alpha$/Fe] relation that
lie on the thick disk sequence according to their position in
the [$\alpha$/Fe]-[Fe/H] plane but lie above the thick or thin disk
sequences in the age-[$\alpha$/Fe] plane. Their age may of course be
underestimated or their enhancements overestimated.  However, there is no
indication of binarity for these stars  nor do there appear to be problems with the determination
of their atmospheric parameters. Despite any worries about outliers, it
is worth noting how remarkably tight the relation age-[$\alpha$/Fe] is for
stars we have classified as part of the ``thin disk'' (ages $<$ 8 Gyr).
At any given age within the thin disk, the dispersion in [$\alpha$/Fe]
is always less than 0.04~dex.  Even more surprising is that, given the
difficulty to derive ages for old stars, the relationship between age and
[$\alpha/Fe$] is still very tight for stars we have classified as ``thick disk''
(i.e., ages $\ga$8~Gyr).
The bottom panel of Fig.~\ref{agealpha2} represents the age-[$\alpha$/Fe] distribution 
with the color coding the absolute magnitude of the stars, showing that the faintest stars 
in our sample introduce no bias in the distribution.

While obviously there are complexities that are underlying this relation,
which we will discuss later, we emphasize that the age-[$\alpha$/Fe]
distribution gives us the opportunity to propose a definition of the
thick and thin disk populations that actually corresponds to two different
epochs of star formation.  These two epochs in the Milky Way appear to
be neatly demarcated at an age of $\sim$ 8 Gyr.  Hence, from now on,
{\it we refer to the thin and thick disk population as stars formed in these
two distinct phases of the Milky Way disk } (solar neighborhood to be more
precise), as manifested in the stellar age-[$\alpha/Fe$] relation.
The appropriateness of this definition when other parameters are taken
into account and how it fits into a more general view of the disk assembly
will be discussed in \S~4 and \S~5.

\begin{figure}
\centering
\includegraphics[width=9.5cm]{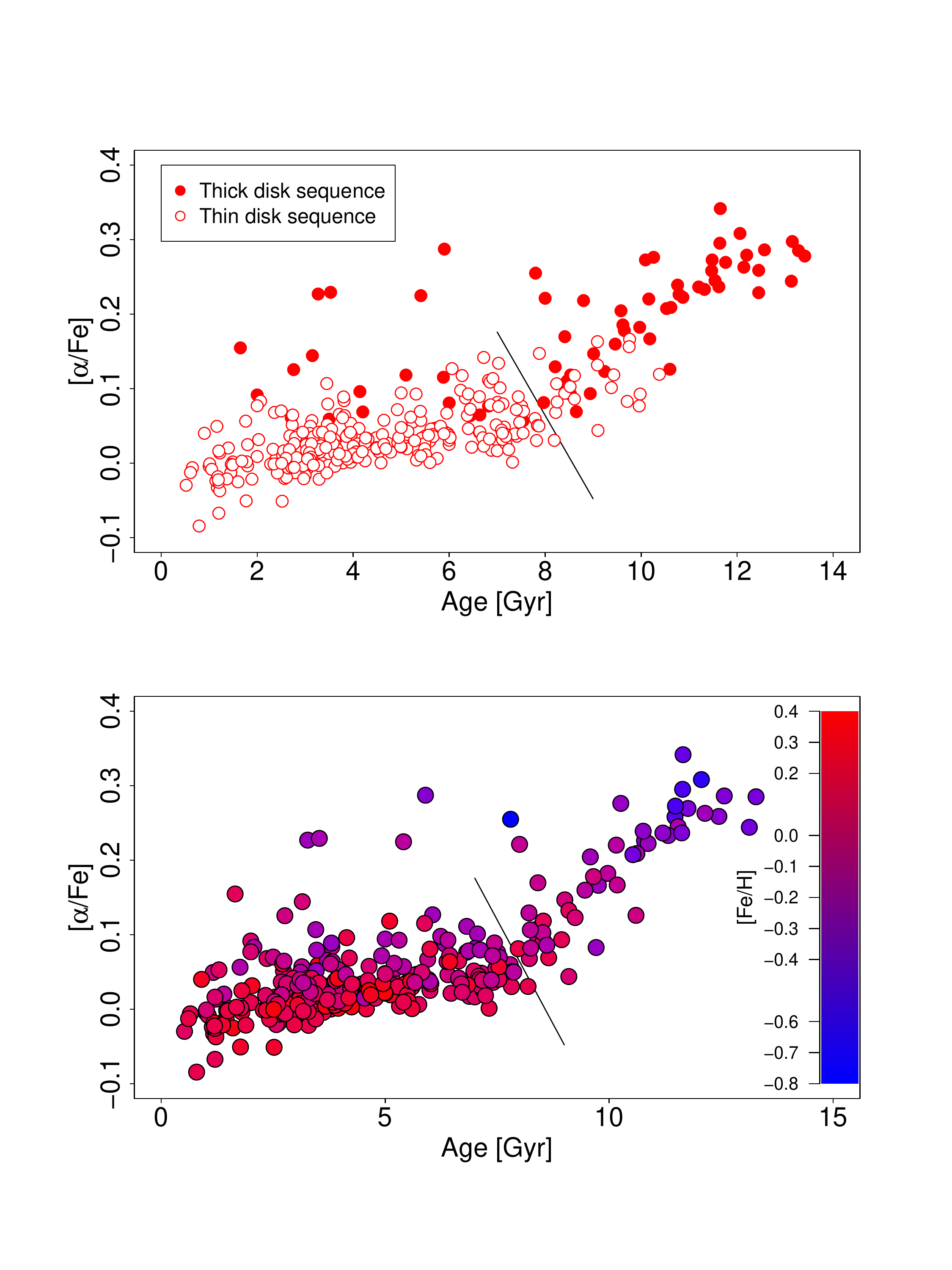}
\caption{{\it Top panel:} 
Same as Fig.~\ref{agealpha2}, but the open
and solid circles represent stars on the thin and thick disks sequence
as classified from their distribution in Fig. 1, respectively. 
{\it Bottom panel:}
Same as above, but color coding based on the metallicity of the stars
as indicated by the color bar.}
\label{agealpha}
\end{figure}

It is first natural to ask how the thin and thick disk stars, often
defined on the basis of their membership to one of the two sequences
in the [$\alpha$/Fe]-[Fe/H] plane (Fig.~\ref{alphafeh}),
redistribute along this new relation.  This is done in the top panel
of Fig.~\ref{agealpha}, where the same stars already present in Fig.~\ref{agealpha2}
are shown, but with symbols now representing objects situated on
the thick disk (solid red circles) and thin disk (open red circles) sequences,
as defined in Fig. \ref{alphafeh}.  Fig \ref{agealpha}, top panel,
shows that: 

\begin{enumerate} 

\item the thin disk sequence, as defined in the $[\alpha/Fe]-[Fe/H]$
plane (open circles), mostly overlaps with the thin disk population -- the
majority of its stars having ages $\leq$ 8 Gyr -- but some stars have
ages between $\sim$8-10 Gyr consistent with the thick disk population.
However,
over the whole time interval from 0 to 10 Gyr, the thin disk sequence
maintains approximately the same $\Delta[\alpha/Fe]/\Delta t$ slope.
We will comment on these important points in the following sections.

\item the thick disk sequence, as defined in the $[\alpha/Fe]-[Fe/H]$ plane,
is mostly coincident with the thick disk population, but some of these
stars have ages younger than 8 Gyr, and, as a consequence, contaminate the
thin disk population. Half of these stars clearly have all characteristics
of thin disk objects and fall into the thick disk sequence only due to
the arbitrary definition of this sequence at [$\alpha$/Fe]$>$0.05 dex.
The other half have above average [$\alpha$/Fe] content (0.05-0.1 dex) at
these high metallicities, and kinematics more akin to thick disk objects.

\end{enumerate}

Bottom panel of Fig.~\ref{agealpha} shows another view of the age-[$\alpha$/Fe] relation with the 
color scale now coding metallicity. Two characteristics are worth noting: the evolution of
the metallicity along the thick disk population, going from about 8~Gyr to 13~Gyr. Second, 
the lower envelope of the distribution in the thin disk regime is occupied by metal-rich stars, while the upper
envelope is occupied mainly by metal-deficient stars of the thin disk (with age $<$ 8~Gyr).

\begin{figure}
\includegraphics[width=9cm]{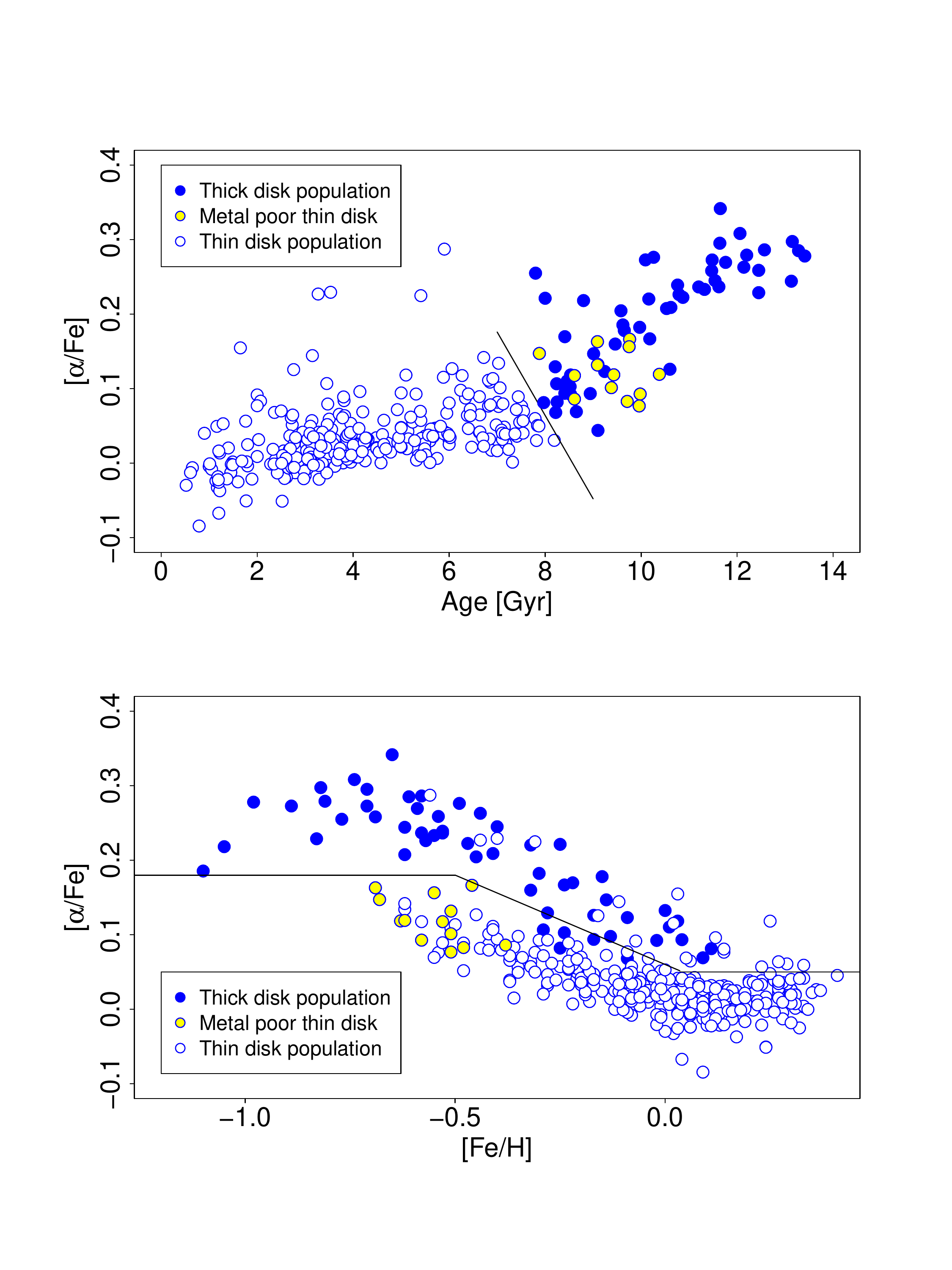}
\caption{{\it Top panel:} [$\alpha$/Fe] $vs$ age for stars in the
sample. The solid black line separates the thin and thick disk populations.
The yellow dots represent the oldest stars that are on the thin disk sequence 
in the  [$\alpha$/Fe]--[Fe/H] plane (see the lower panel), and for which age could be determined.  
These objects fall in the part of the age--[$\alpha$/Fe] distribution that corresponds to the thick 
disk (above the line).
{\it Bottom panel:} [$\alpha$/Fe]
$vs$ [Fe/H] for the same sample and with the symbols indicating the same
population of stars as above. This illustrates how the two sequences are separated
in this plane.}
\label{agealpha1}
\end{figure}

Conversely, it is interesting to see how stars in the two segments
of the age-[$\alpha$/Fe] relation are distributed within the
[$\alpha$/Fe]--[Fe/H] plane.  This is shown in Fig.~\ref{agealpha1}, which
shows (top panel) the two thick and thin disk populations separated by
a (somewhat arbitrary) diagonal line. The yellow symbols represent the
old metal-poor thin disk stars which mix with thick disk stars in the
[$\alpha$/Fe]--[Fe/H] plane (shown in the bottom panel).  We conclude
that:

\begin{enumerate} 

\item  Given the remaining uncertainties in the age determination,
and the fact that there is some arbitrariness in the positioning of the
separating line -- and apart from the specific problem of the metal-poor
thin disk stars which we address below -- our classification for the
age--[$\alpha$/Fe] plane corresponds to a remarkable separation of the
two populations in the [$\alpha$/Fe]--[Fe/H] plane.

\item The oldest metal-poor thin disk stars are responsible for the contamination of the thick disk population by thin disk objects in the age-[$\alpha$/Fe] distribution.
 Haywood (2008) proposed that thin disk stars with metallicities below about $\sim$-0.3~dex (of almost all ages: as young as 2 Gyr and up to the stars
discuss here with ages $>$ 8~Gyr, see Fig.~\ref{agemet})
originate from the outer thin disk.  We contend that they have no
direct connection with the thick disk (but see section 5.2), despite having similar ages and
[$\alpha$/Fe] content, neither are they progenitors of the younger (age
$<$8 Gyr) inner disk stars. 
The following arguments favors an outer disk origin for these objects. 
Several of these stars have apocenters reaching galactocentric distances 
larger than 9 kpc, being the only disk population in the solar vicinity showing this 
characteristic. This was demonstrated in Haywood (2008) for stars in the solar vicinity, and is also clearly 
visible in Fig. 7 of Bovy et al. (2012b) for a more spatially extended sample. 
While these metal-poor stars represent only a few percent of the local disk, 
objects with similar chemical properties dominate the disk at R$>$9-10 kpc, or  
about
1-2 kpc outside the solar orbit. With a dispersion in the U-component of 
about 50 km.s$^{-1}$, one derives an epicycle radial excursion of 1-2~kpc (e.g, Ro{\v s}kar et al. 2011), 
which is sufficient to explain that some of these objects contaminate local samples.
Further arguments and a discussion of the status of these stars and how they fit in our
general scheme are presented in \S~5.1.

\end{enumerate}

\subsection{The age-metallicity relation}\label{individual2}

\begin{figure}
\includegraphics[width=9cm]{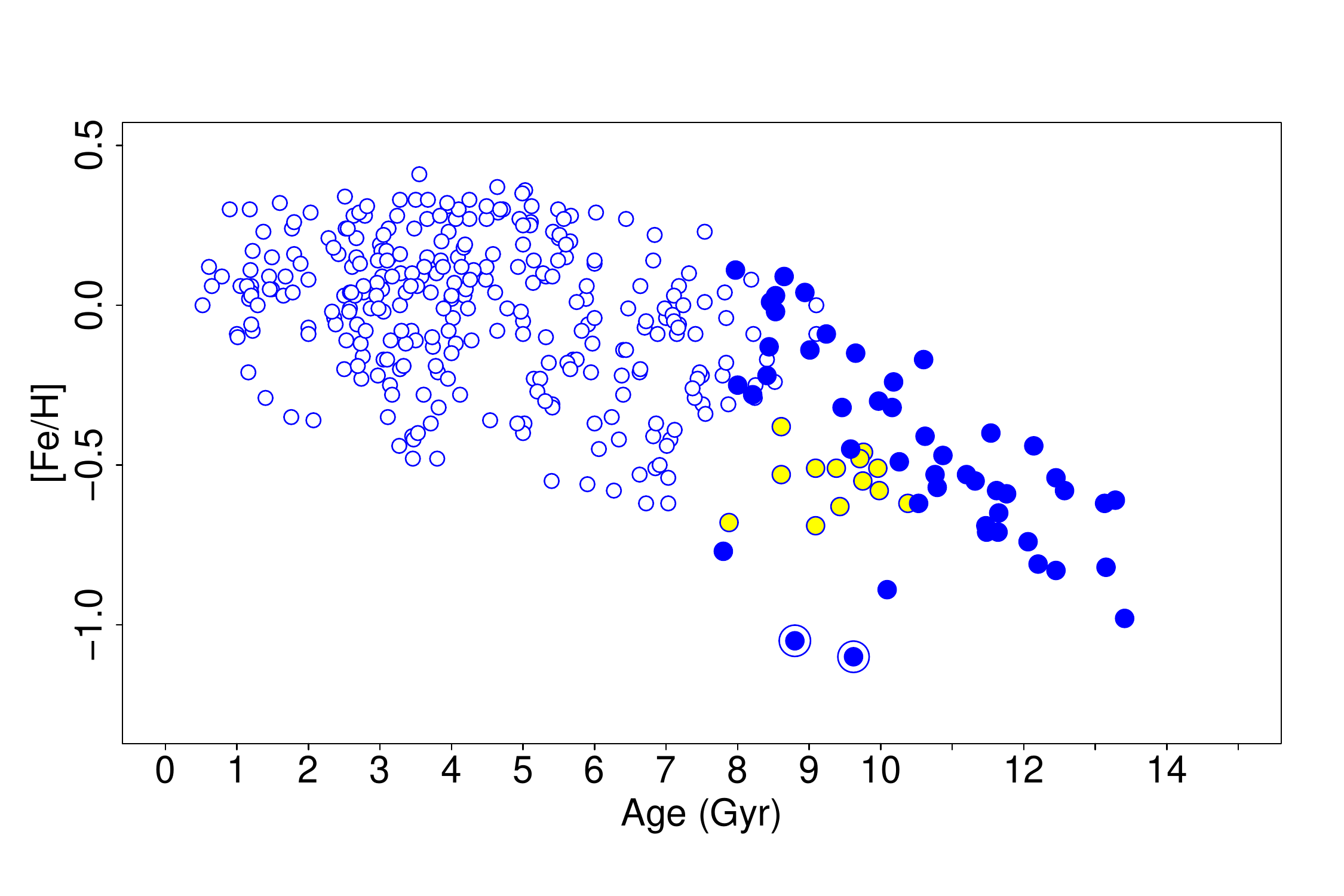}
\caption{Metallicity as a function of age for stars in the sample.
Symbols as in the top panel of Fig.~\ref{agealpha1}. When thin disk
metal-poor stars (yellow dots) are not considered, the
age-metallicity relation of thick disk stars (blue solid circles)
is obvious. The two objects at [Fe/H]$<$-1.0 dex and age $\sim$ 9 Gyr are 
HIP 54641 and HIP 57360. See the text for details.}
\label{agemet}
\end{figure}

In light of the insights gained in investigating the relationship
between age and $\alpha$-enhancement, it is interesting to look at
the age-metallicity ([Fe/H]) relation.  The age-metallicity relation
confirms that stars in the thick disk show a much tighter correlation
between age and metallicity than thin disk stars (Fig.~\ref{agemet}).
Moreover, the increase in metallicity in the thick disk phase
($\sim$0.15~dex~Gyr$^{-1}$) is much steeper than in the thin disk
(0.025~dex~Gyr$^{-1}$), also implying a decrease by a factor of 5--6 in
the production of iron after 8 Gyr.  Moreover, metal-poor thin disk stars
are not degenerate in the age-[Fe/H] as they are in the [$\alpha$/Fe]--age
plane. Once they are identified in the age-metallicity distribution,
the correlation between age and metallicity for thick disk stars becomes
much clearer.  Note the four outliers to the thick disk age-metallicity
relation.  Two of them have no particular characteristics,
and seem to be standard thick disk objects. We can offer no particular
explanation for their `young' ages.  The two others (circled symbols on
Fig. \ref{agemet}) are HIP~54641 and HIP~57360, two stars with slightly
low alpha abundances (see Fig. \ref{alphafeh}), and significant U velocity
(respectively +84 and +100 km s$^{-1}$) . Taken together, these arguments
suggest that these two stars have been accreted.

At ages less than about 8 Gyr, in the thin disk regime, the dispersion
in metallicity increases sharply.  This is in agreement with a number of
previous studies \citep[e.g.][]{edv93, nor04, hay06}. 
This may be mainly an effect of radial migration of
the stars through ``churning'' or ``blurring''\footnote{Following the terminology introduced by Sch{\"o}nrich \& Binney (2009a), we 
use the term ``churning'' to characterise a change of guiding center with
no kinematic heating, as proposed by \citet{sel02}, and ``blurring''
for the increase in eccentricity, or epicycle radial oscillation of the orbit. }.
However,
it is unclear which type of mixing is responsible for the dispersion in
metallicity measured in the solar vicinity, because it is known that the
Sun is near the dividing line between the inner disk, which is enriched
in metals (mean [Fe/H]$\sim$+0.2 dex at R$<$7kpc, see Hill et al. 2012),
and the outer disk which is depleted in metals (mean [Fe/H]$\sim$$-$0.3
dex at R$>$10 kpc, field stars: \citet{ben11} and reference therein,
open clusters: \citet{jac11} and reference therein).  As a result of the Sun lying near this dividing line, it is
possible that the solar neighborhood has contamination from stars with
a relatively wide range of metallicities but which did not originate
very far from the solar orbit.  We comment further on this point and
the importance of radial migration in the Milky Way in \S~3.5.2 and \S~5.2.

\begin{figure}
\includegraphics[trim=40 0 0 0,clip,width=10.cm]{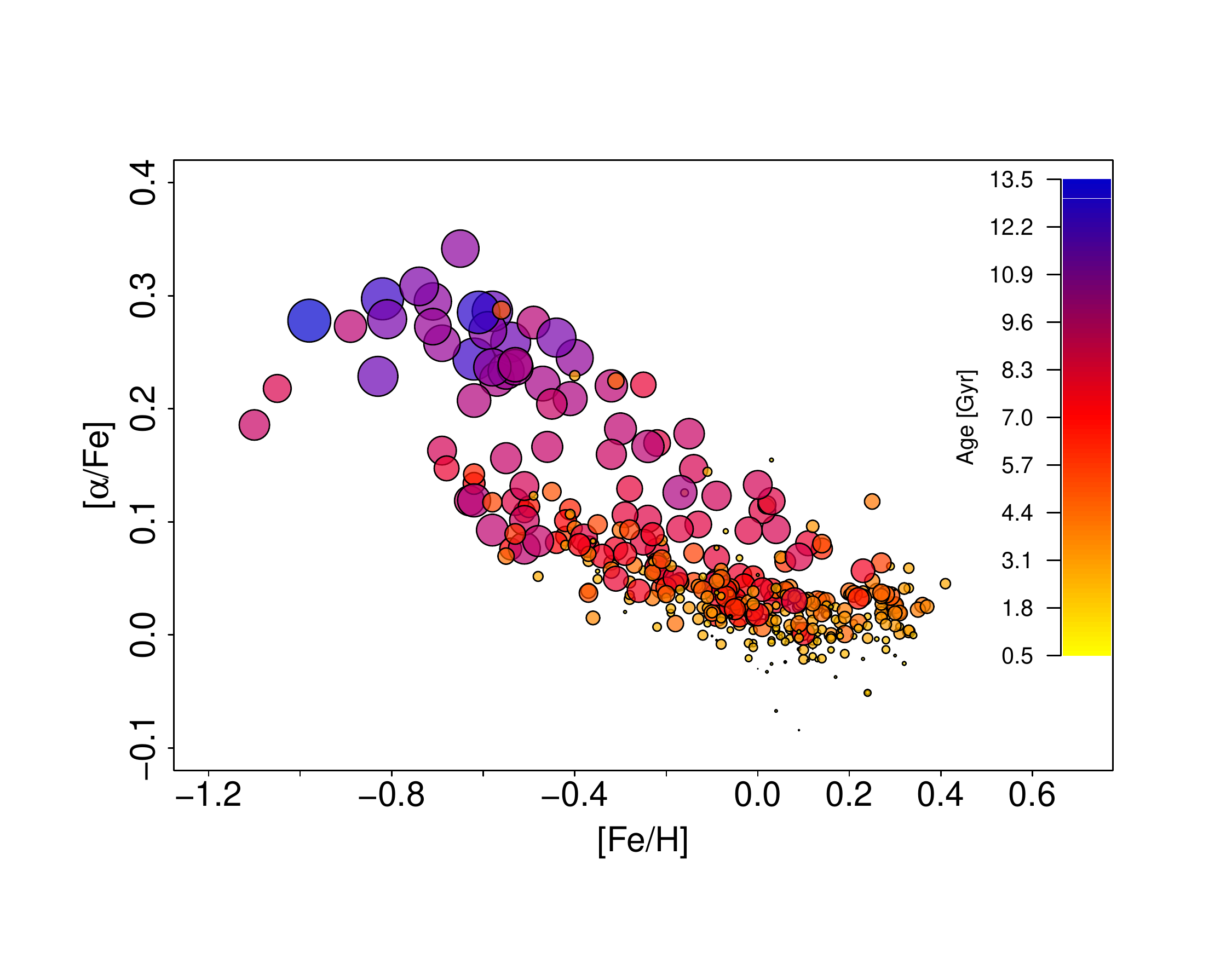}
\caption{ [$\alpha$/Fe] $vs$ [Fe/H] for the stars in the sample of
Adibekyan et al. for which a robust age could be derived. The color and
the size of the symbols both code the age of the stars, to emphasize
the age stratification of the distribution of stars within this plane.  }
\label{alphafehage}
\end{figure}

\subsection{The age structure within the $[\alpha/Fe]$--$[Fe/H]$ distribution}

We have investigated the [$\alpha$/Fe]--age and age--metallicity
relations separately, but does there exist a temporal sequence within
the [$\alpha$/Fe]--[Fe/H] plane?  Any possible relationship is shown in
Fig. \ref{alphafehage}, where the [$\alpha$/Fe]-- [Fe/H] distribution
is shown for those stars in the sample which have well-determined ages.
The main findings from this investigation are:

\begin{enumerate}

\item There is a clear tendency of ages to decrease with [$\alpha$/Fe],
no matter what the metallicity.  

\item For the thick disk population, because of the tight relation
between metallicity and age, this tendency reflects into a clear
temporal sequence: the thick disk evolved with time from a metal
poor ([Fe/H]$\sim$$-$0.8 dex), $\alpha$-enriched population with ages
$\sim$13~Gyr, to a younger  (ages$\sim$8~Gyr), metal-rich population with metallicities around solar and an mild
$\alpha$-enhancement ($[\alpha/Fe]\sim$0.1~dex).
\citet{adi11, adi13} have discussed the status of the most metal-rich 
stars at the end of the thick disk sequence as being a mix of thin and thick disk 
stars. With the new definition of thin and thick disks introduced in section \ref{individual}, 
most objects are classified as thin disk. Out of 28 stars with [Fe/H]$>$-0.1 dex, [$\alpha$/Fe]$>$0.05 dex
and for which age has been obtained, 9 are in the thick disk regime.

\item  Unsurprisingly, within the thin disk, the age sequence becomes
much less evident. There is still a tendency for stellar population to
move with decreasing age from lower metallicities and higher $\alpha$
abundances towards higher metallicities and solar and sub-solar $\alpha$
abundances. However, as already known, the dispersion in [Fe/H] is large,
whatever the [$\alpha$/Fe] or age. However, the metallicity dispersion
is expected to increase with the relative alpha element abundance as it
does with age, due to dynamical effects which progressively radially mix
stars in the disk.  This is confirmed in our study of this sample: the
dispersion in the metallicity increases with increasing age from about
0.14~dex at [$\alpha$/Fe]$<$0.0~dex to 0.25~dex at [$\alpha$/Fe]=0.1~dex.

\end{enumerate}

\subsection{Kinematics}

\subsubsection{Age-vertical velocity dispersions}

\begin{figure*}
\includegraphics[width=19cm]{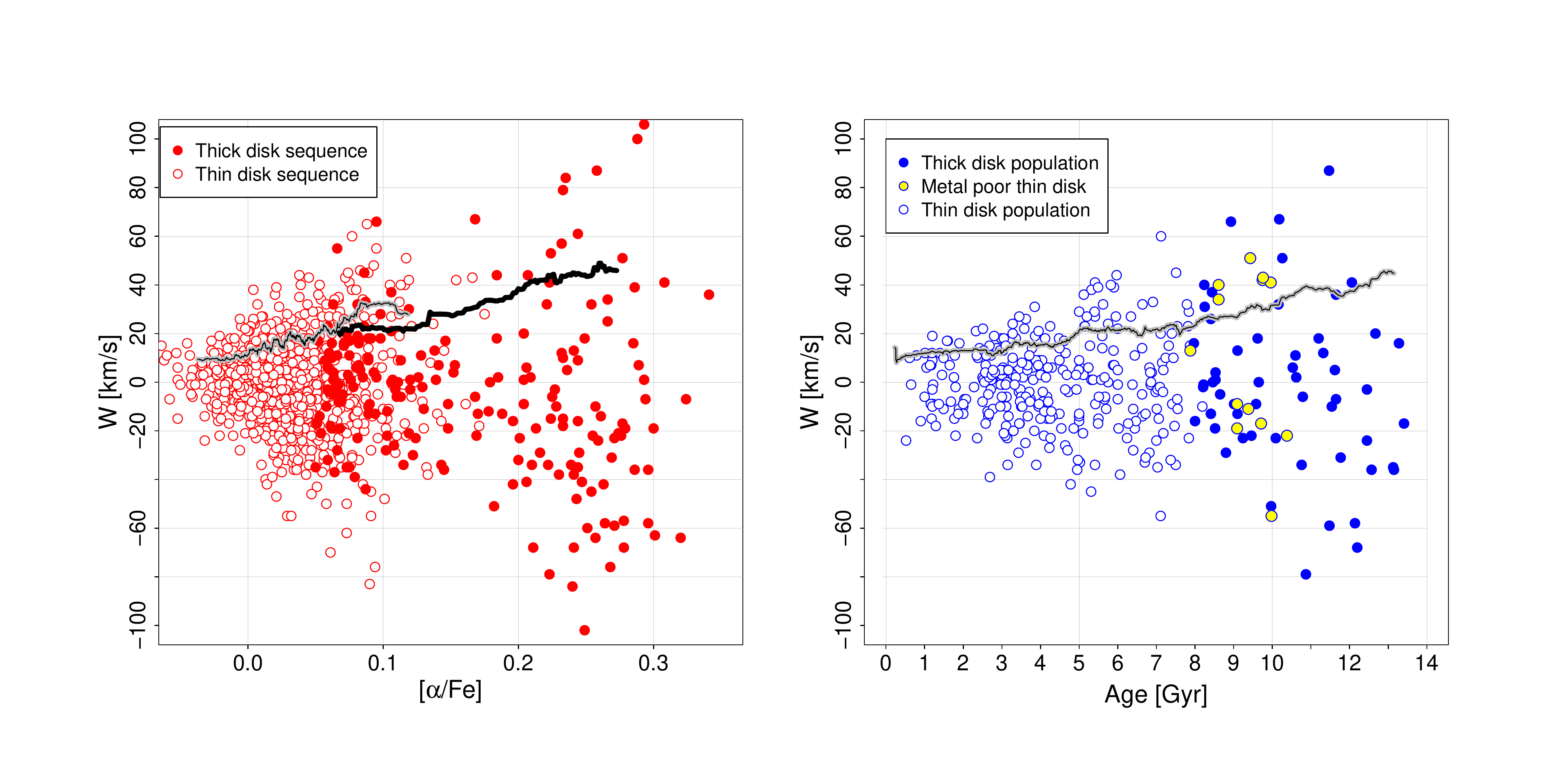}
\caption{
{\it Left panel:} W component of velocity as a function of
[$\alpha$/Fe] for the stars in the sample (open red circles represent
stars on the thin disk sequence; filled red circles represent stars on the
thick disk sequence).  The two thin and thick black curves are running
dispersions for W component of the velocity of the thin and thick disk
sequence stars, calculated as a function of [$\alpha$/Fe] (over 50 stars).  {\it Right panel:} W component of velocity as a
function of age. Open blue circles represent thin disk stars and solid
blue circles represent thick disk stars as defined on Fig. 5. The curve is 
a running dispersion calculated on 50 stars.
}
\label{wsigalpha}
\end{figure*}

Fig. \ref{wsigalpha} shows the W velocity component as a function of
[$\alpha$/Fe] and age, for the stars in the sample.  The two curves in
each panel are the running dispersion in the W velocity calculated over
50 points.  The dispersion along the thin
disk sequence increases from about 9$\pm$1.5 km s$^{-1}$ to 35$\pm$6 km
s$^{-1}$, while in the thick disk sequence, the dispersion varies from
22$\pm$3.7 km s$^{-1}$ to 50$\pm$8.3~km s$^{-1}$. It is interesting to
note that it is the group of old, metal-poor thin disk stars that is
responsible for the higher dispersion of 35~km~s$^{-1}$ in the thin disk
sequence. When these objects are discarded by selecting thin disk sequence
stars with $[Fe/H]$>$-$0.3~dex ($-$0.4~dex), the vertical dispersion rises
to only 22~km~s$^{-1}$ (27~km~s$^{-1}$).  This is confirmed by the data
shown in the right panel, where the vertical dispersion rises to about
the same values at age$\sim$8~Gyr.

\begin{figure}
\includegraphics[width=9cm]{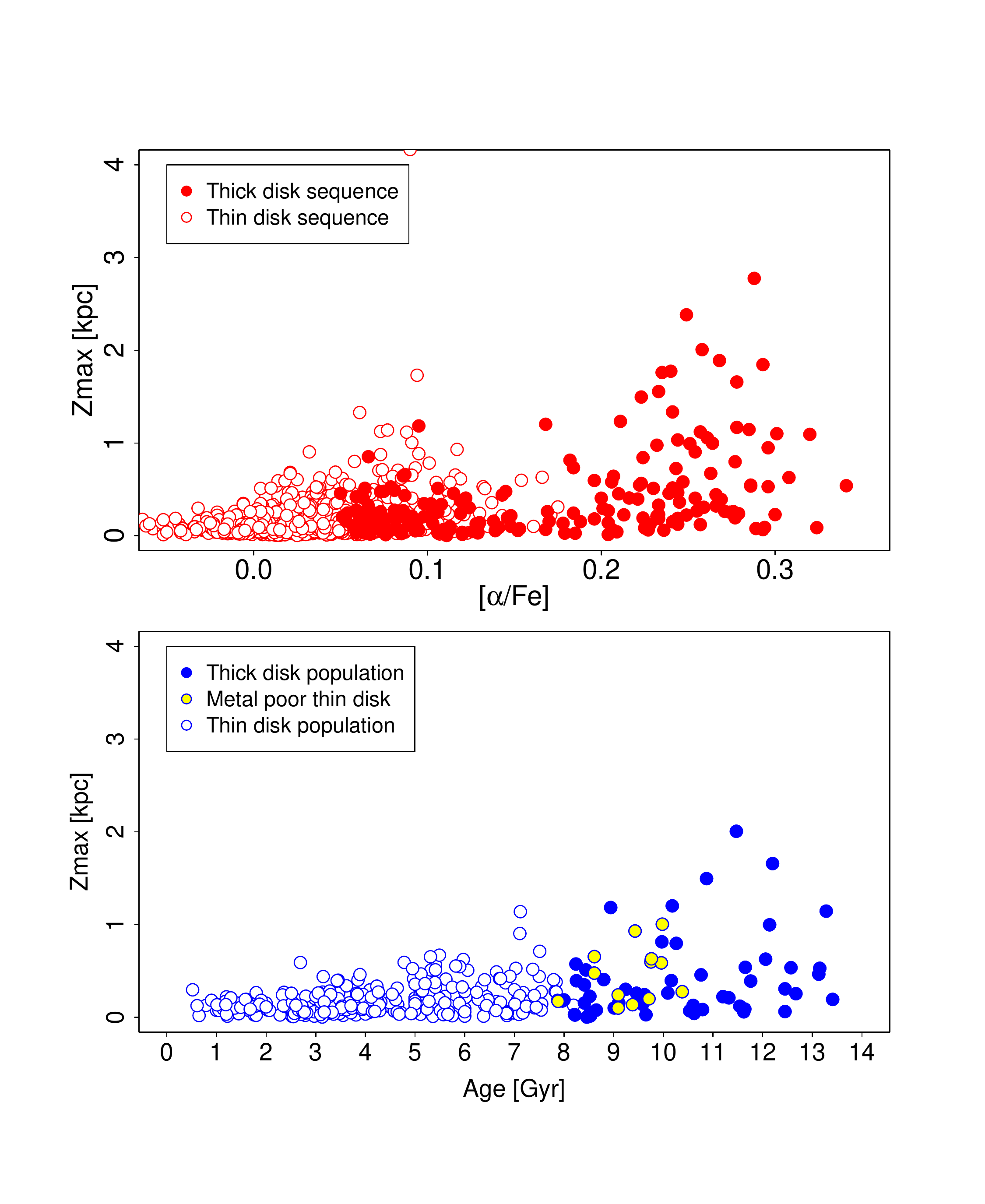}
\caption{Maximum distance from the galactic plane as function of
[$\alpha$/Fe] abundance (upper plot) and age (lower plot). Open (red
or blue) circles represent thin disk stars and solid (red or blue)
circles represent thick disk stars as defined on Fig. 5.}
\label{zmaxage}
\end{figure}

The $z_{max}$ values corresponding to the vertical velocities are plotted as a function
of [$\alpha$/Fe] and age in Fig.~\ref{zmaxage}, clearly illustrating the decrease
in scale height within the thick disk population.

\begin{figure*}
\includegraphics[width=19cm]{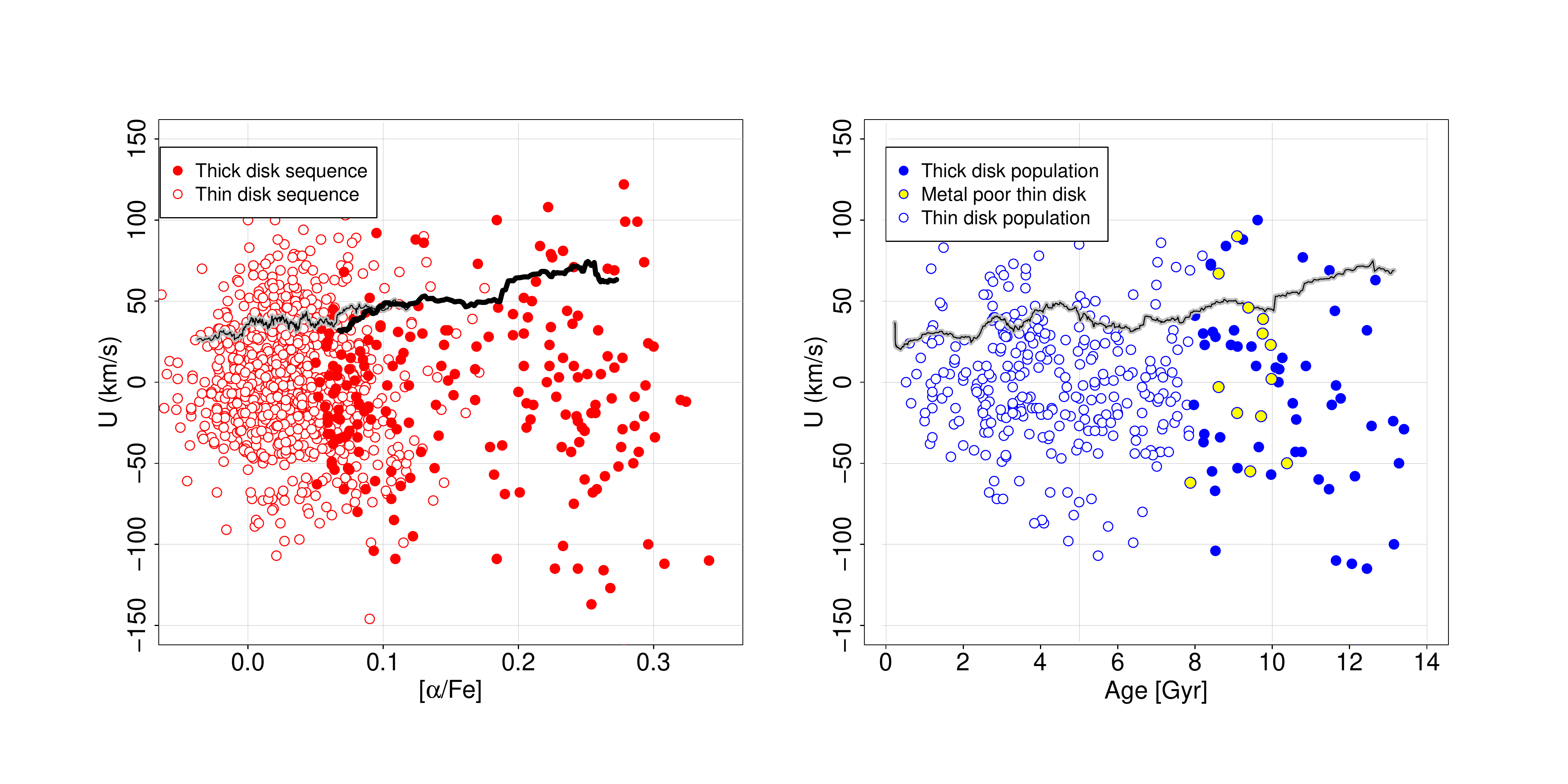}
\caption{
{\it Left panel:} U component of velocity as a function of
[$\alpha$/Fe] for the stars in the sample (open red circles represent
stars on the thin disk sequence; filled red circles represent stars on the
thick disk sequence).  The two thin and thick  curves are running
dispersions for U component of the velocity of the thin and thick disk
sequence stars, calculated as a function of [$\alpha$/Fe] (over 50 stars).  {\it Right panel:} U component of velocity as a
function of age. Open blue circles represent thin disk stars and solid
blue circles represent thick disk stars as defined on Fig. 5.
}
\label{usigalpha}
\end{figure*}

The so-called age-$\sigma_W$ relation,
which has been investigated intensely in the hope of measuring a
saturation value or a step that would indicate a transition from the
thick to the thin disk, mixes stars of different provenance and which for
a given age, have different vertical dispersions. In mixing stars of
different provenance, finding a transition may be spurious. We note
that stars along the thin disk population with ages of $\sim$8~Gyr have
similar dispersion as stars along the thick disk sequence with ages of
$\sim$9-10~Gyr, probably due to the same process of vertical heating.
Paradoxically, stars of the metal-poor
thin disk, also being 9-10 Gyr old, have a dispersion higher than that of
the thick disk of the same age.  Therefore, we should not be surprised
that samples comprising different amount of metal-poor thin disk,
``young'' thick disk, and old thin disk, would produce different overall $\sigma_W$ at a
given age, being a mix of stars of different populations with different
vertical energies. We emphasize that discussing an age-$\sigma_W$ relation
is meaningful only if the contributions of the different components are
properly disentangled.

The fact that metal-poor thin disk stars have higher vertical velocity
dispersions, together with their probable outer disk origin, suggests the
intervention of some dynamical mechanism operating in the outskirts of
the disk that may add some extra vertical kinetic energy.  A warp could
produce an increase in the velocity dispersion. Specific signatures, as an
asymmetry in the distribution of vertical velocities, have been searched on
local data, with contradictory results \citep{deh98,sea07}. 
The mean of the vertical velocities of our metal-deficient thin
disk stars is 2~km~s$^{-1}$, compatible with no asymmetry.  Flat profiles
of the vertical velocity dispersion in external parts of spirals have
been found in a number of objects \citep{her09}, and a 
flaring of the disk has been suggested to be responsible for such effect.

Fig. \ref{usigalpha} shows the U velocity component as a function of age and [$\alpha$/Fe]. 
The left panel shows that the velocity dispersion reaches 50 km s$^{-1}$ for the
metal-poor (alpha-rich) thin disk stars, significantly higher than the standard thin 
disk (30-40~km s$^{-1}$). Overall, the outer thin disk appears to be significantly hotter 
than the old, local, thin disk, with values not far from stars of intermediate age 
in the thick disk,  while maintaining a substantial rotation velocity. 
By virtue of the asymmetric drift relation, this implies that the scale length of
metal-poor thin disk stars must be substantially larger than the one of the inner thin 
disk. In the disk decomposition made by \citet{bov12b} using the SEGUE data, it is 
noticeable that the low-metallicity, low-alpha abundance component, 
has the longest scale length measured of all thin disk components, 
with h$_R$=4.3 kpc. By comparison, the most metal-rich components have scale lengths
of the order of 2.3-2.8 kpc.

\subsubsection{Rotational velocities versus [Fe/H], [$\alpha$/Fe] and ages}

\begin{figure}
\center
\includegraphics[trim=100 0 0 0,clip, width=11.cm]{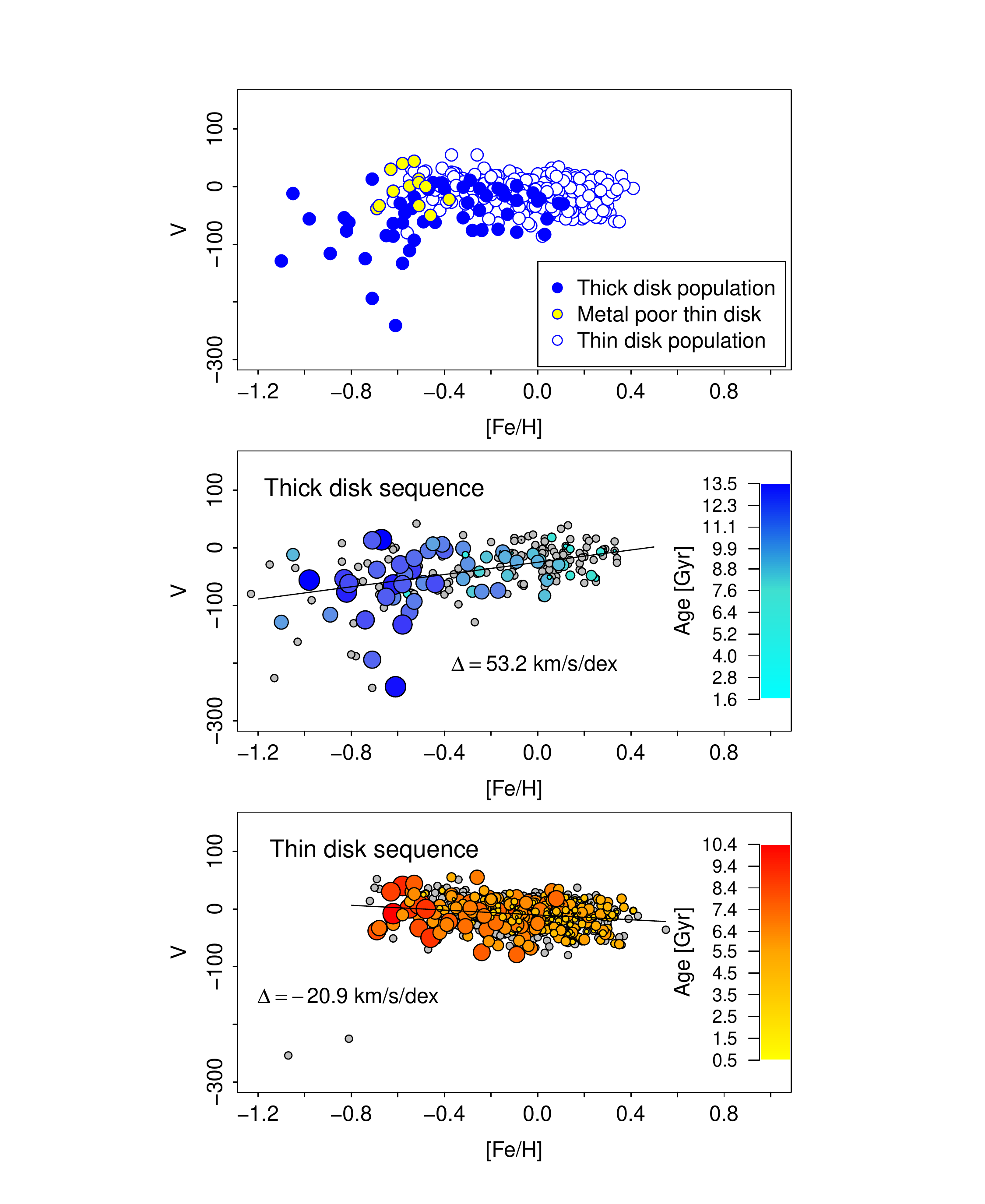}
\caption{
\emph{Top panel:} 
V component of the velocity as a function of [Fe/H] for the thick disk (filled circles)
and thin disk (open circles) populations, as defined in Fig. \ref{agealpha2}.
\emph{Middle panel:} 
V component of the velocity as a function
of [Fe/H] for thick disk sequence stars in the sample.
\emph{Bottom panel:} 
V component of the velocity as a function
of [Fe/H] for thin disk sequence stars in the sample.
In the middle and bottom panels, stars without age determinations are shown as solid grey
circles.  For stars with age determinations, the color and
the size of the symbols indicate the age using the scale given in the color bars
shown on the right of each plot.
Solid black lines represent the best fit linear regression to the data, with the fit coefficient  indicated. For the thick disk sequence, only stars in prograde orbits (V > $-$220~km~s$^{-1}$) have been included in the fit. For thin disk
sequence stars, the two points with V < $-$200~km~s$^{-1}$ have not been
included when evaluating the best fit linear regression of the data.}
\label{metV}
\end{figure}

\begin{figure}
\includegraphics[trim=100 0 0 0,clip, width=11.cm]{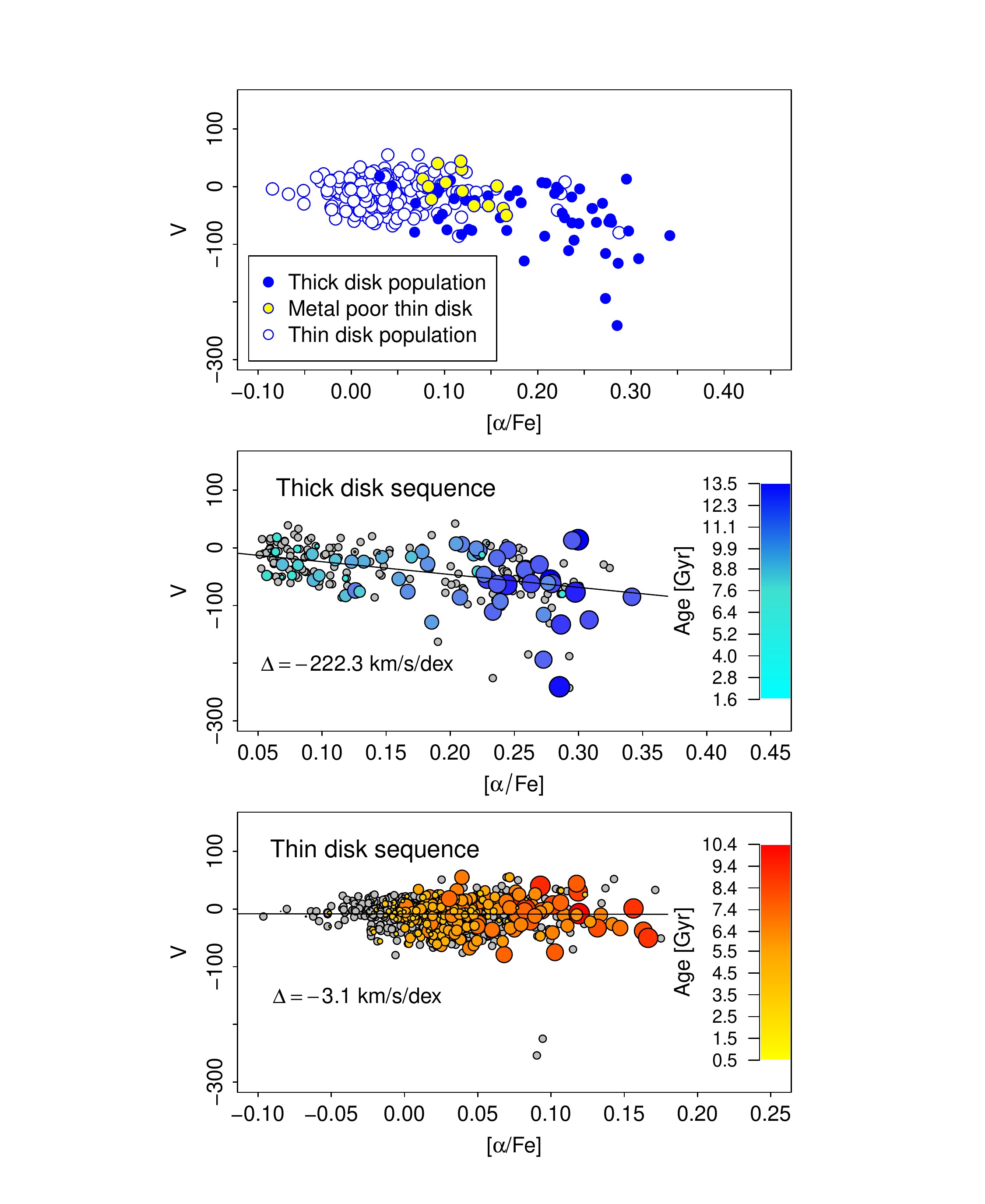}
\caption{
\emph{Top panel:} 
V component of the velocity as a function of [$\alpha$/Fe] for the thick disk (filled circles)
and thin disk  (open circles) populations, as defined in Fig. \ref{agealpha2}. 
\emph{Middle panel:} 
V component of the velocity as a function of
[$\alpha$Fe] for thick disk sequence stars in the sample.
\emph{Bottom panel:} V component of the velocity as a function of
[$\alpha$Fe] for thin disk sequence stars in the sample.
In the middle and bottom panels, stars without age determinations are shown as solid grey
circles.  For stars with age determinations, the color and
the size of the symbols indicate the age using the scale given in the color bars
shown on the right within each plot.
Solid black lines represent the best fit linear regression to the data, with the fit coefficient  indicated. For the thick disk sequence, only stars in prograde orbits (V > $-$220~km~s$^{-1}$) have been included in the fit. For thin disk
sequence stars, the two points with V < $-$200~km~s$^{-1}$ have not been
included when evaluating the best fit linear regression of the data.}
\label{alphaV}
\end{figure}

\begin{figure}
\includegraphics[trim=100 0 0 0,clip, width=11.cm]{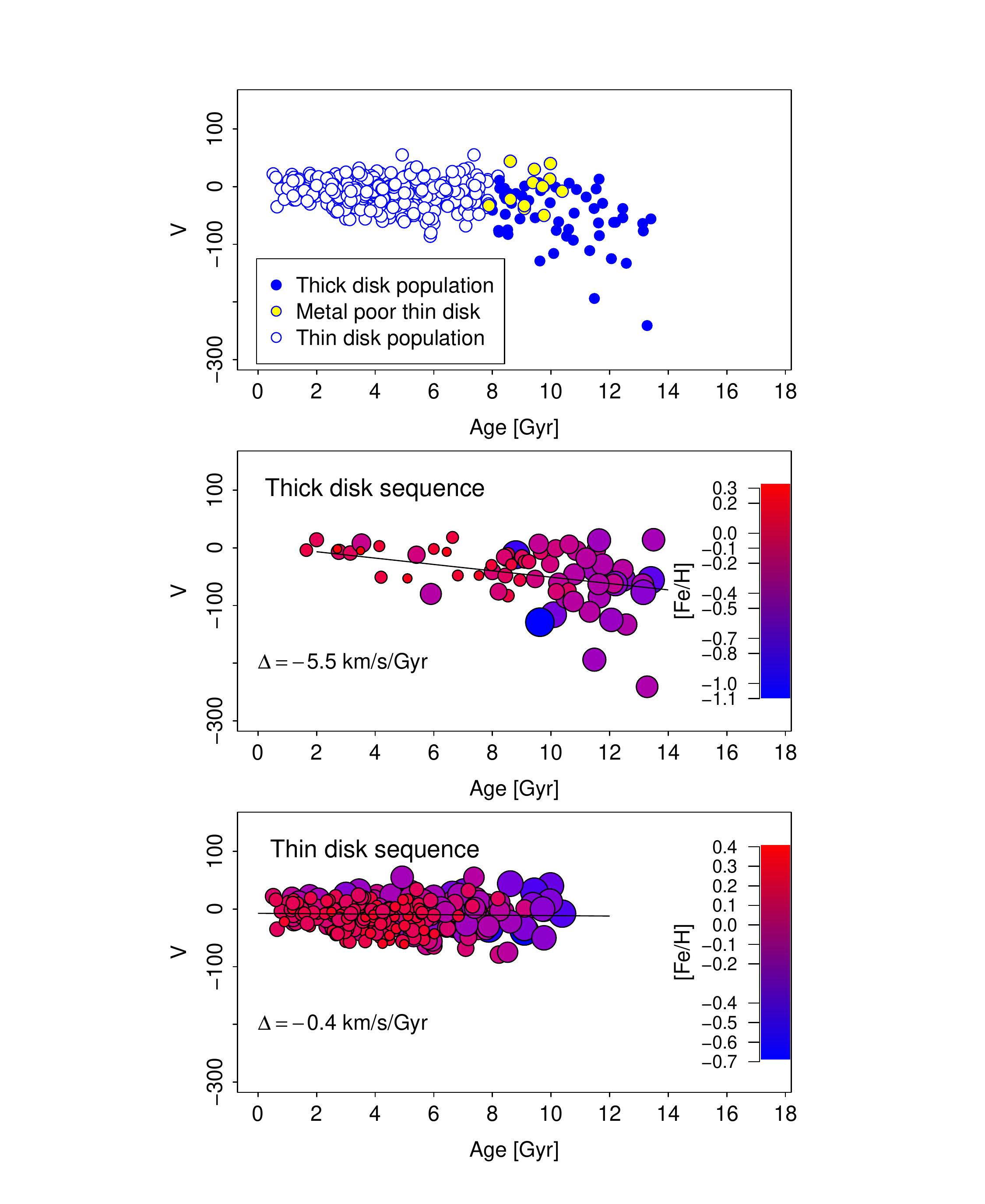}
\caption{
\emph{Top panel:} V component of the velocity as a function of
age for thick and thin disks population stars, as defined in Fig.~\ref{agealpha2}.
\emph{Middle panel:}
V component of the velocity as a function of
age for thick disk sequence stars. \emph{Bottom panel: }V component of the
velocity as a function of age for thin disk sequence stars.  In all the
panels,  the color and the size of the symbols are a function of a stars
[Fe/H] value as indicated by the color bar in each plot. Solid lines
represent the best fit linear regression to the data. For the thick disk sequence, only stars in prograde orbits (V > $-$220~km~s$^{-1}$) have been included in the fit. For thin disk
sequence stars, the two points with V < $-$200~km~s$^{-1}$ have not been
included when evaluating the best fit linear regression of the data.}
\label{ageV}
\end{figure}

The  relation between the rotational support, V, of stars and their
chemical properties has been extensively studied as a way to measure
the Galactic disk evolution.  However, the introduction of the concept of
radial migration in the last years has lead to conflicting results
both  in  the predictions  of  models and in the data analysis and
interpretation \citep[see, for example,][]{ive08, sch09b, spa10, nav11, loe11, lee11, cur12}.  
\citet{loe11} claimed that such correlation should be  erased in the presence of substantial
mixing. Note,  however,  that  starting from rather different, opposing assumptions,
\citet{nav11} interpreted  the absence of a V--[Fe/H] relation
in the local  thin disk as an absence of  migration processes.  On the
contrary,  \citet{sch09b},   suggested   that  this
correlation may  subsist even in  the presence of  substantial mixing,
because stars coming  from the inner disk will  still dominate the low
angular  momentum  values  even  though  they  have  migrated  through
churning.  
In  their picture,  it  is  the
strength (or weakness) of the observed V--[Fe/H] relation at the solar
vicinity which constrains the amount of churning in the disk. If stars
did not experience any angular momentum change (e.g. mixing only by
blurring), the conservation of angular momentum for stars coming 
from several kpc - as suggested by their extreme metallicities, and in the hypothesis
of a linear gradient -  would imply extreme
rotational  velocities that  are not  seen in  the solar  vicinity
metallicity gradient. 
We shall  discuss   actual  evidence  of  radial   migration  in  \S~\ref{migration}.

There is also a lot of confusion as to how samples are defined and interpreted. 
\cite{loe11}, for example, pointed out the absence of a V-[Fe/H] correlation 
among old disk stars in the Geneva-Copenhagen Survey sample \citep[GCS,][]{hol09}.
But \citet{loe11} use undistinguished old
disk stars, without classifying them as either belonging to the thin and thick disk populations. 
As pointed out by \citet{hay12}, the mix of lagging thick disk stars and fast rotating metal-poor thin disk stars could erase a correlation in the thin disk. \citet{nav11} concluded also on the
absence of such a correlation among thin disk stars, 
defining a sample using a single limit in [$\alpha$/Fe] over the whole range of metallicities, but which has the effect of mixing the thick and thin disks.
These results seems in contradiction with the results of \citet{spa10}, and \citet{lee11}, who both reported the existence of a 
rotational velocity-metallicity gradient in the thick disk. 

In this context, we have analyzed the trends in  rotational velocity
versus metallicity for the stars in our sample.  The top panel in the Figure~\ref{metV} shows the V-[Fe/H] relation for stars in the thin and thick disk population. We remind the reader 
that since the membership of stars to one of these two populations is defined according to their location in the  [$\alpha$/Fe]-age relation (Fig.~\ref{agealpha2}), 
in this panel only a subset of the whole sample is shown. The data show a flat trends for stars of the thin disk population, with a slight decrease of the 
rotational velocity only for the less metal poor thin disk stars ([Fe/H]$<$-0.5~dex). Stars belonging to the thick disk population show a more complex trend, 
characterized by the absence of a V-[Fe/H] gradient for [Fe/H]$>$-0.5 and a positive  V-[Fe/H] correlation for more metal poor stars  ([Fe/H]$<$-0.5~dex). 
Overall the two populations show a continuity in the rotational velocity-metallicity trend, with thick disk stars at [Fe/H]$>$-0.5~dex having a slightly 
lower rotational support than thin disk stars of the same metallicity. When separating stars in the thick and thin sequences (as defined in Fig.~\ref{alphafeh}), 
rather than in the two populations, similar trends are found. Thick disk stars  show an overall increase in the rotational velocity with metallicity, 
the gradient of the relation being $\Delta_{V-Fe/H}=53.9$~km/s/dex (central panel, Fig.~\ref{metV}). To evaluate the gradient all the stars of the thick
disk sequence have been included, e.g. those with ages estimates (colored point in the plot) and those without (grey points).  Only few stars in retrograde 
motion (V$< -$220 km/s) have been excluded from the linear fit. Note however that while the overall sample of thick disk sequence stars can be represented 
by a single linear fit, we notice the same tendency already found for the thick disk population, i.e. a flattening of the relation for [Fe/H]$>$-0.5~dex. 
This  trend, already noticed by \citet{spa10} for metallicity as high as  [Fe/H]$=$-0.3~dex, is maintained all along the metal rich thick disk sequence, 
up to supersolar metallicities. The existence of a double gradient in the V-[Fe/H] relation for thick disk stars may reconcile the apparent contradictory 
results, as those by \citet{lee11} and \citet{loe11}. Indeed, the first focused their analysis on thick disk stars with metallicities lower than  
[Fe/H]$= -$0.5--$-$0.3~dex, while the second evaluated the gradient for old disk stars, with Fe/H]
$>$-0.5~dex, 
but note that with this selection, thick disk stars must have remained a minority.\\

Figs.~\ref{alphaV}-\ref{ageV} show similar trends: 
\begin{itemize}
\item$\alpha$-enriched, old stars tend to rotate slower than $\alpha$-poor, young stars;
\item there is a continuity in the evolution of V with $\alpha$ elements and ages;
\item the thick disk sequence shows a positive correlation between V and $\alpha$ (or age), while the thin disk sequence shows a flat gradient;
\item the evolution of the rotational support with age proceeds more than ten times faster among thick disk sequence stars ($\Delta_{V-age}=-5.5$~km/s/Gyr) than for the thin disk counterpart ($\Delta_{V-age}=-0.4$~km/s/Gyr).
\end{itemize}

\section{Implications} \label{imp}

Before discussing the implications, it is worthwhile to summarize the
results from the previous section. We enumerate our results as follows:

(1) There are two distinct regimes in the decline of [$\alpha$/Fe] and
the increase in [Fe/H] with age. One regime, at early epochs, shows [$\alpha$/Fe] decreasing
5 times faster and also [Fe/H] increasing more than
10 times faster than in the second regime, at later times.  The two regimes separate rather
cleanly at ages around 8 Gyr.

(2) The age-metallicity and age-[$\alpha/Fe]$ correlations in the thick
disk regime have small scatter.

(3) The vertical velocity dispersion decreases with decreasing
$[\alpha/Fe]$ (or age) for stars along the thick disk sequence.

(4) The oldest stars in the thin disk have metallicities around $-$0.6~dex
and  ages$\approx$10~Gyr.  Although these objects have similar age and
[$\alpha$/Fe] content as the youngest thick disk stars, they have very
different metallicities and have a higher rotational component in their
velocities. Both these points can be considered as indicating that these stars
originated in the outer disk \citep{hay08}.

Point (1) provides evidence for the existence of two different epochs
of star formation in the Galaxy, which we have defined as the epochs
of thick disk and thin disk formation. The change between these two
epochs is well segregated in time, and suggests that the conditions
under which stars formed were significantly different before and after
$\sim$8 Gyr. However, this should not be taken to imply that there was a
discontinuity between the formation of the two disks.  The marked change
of slope in the age-metallicity relation shows that it is the decrease
in the production of iron at $\sim$8 Gyr that is mainly responsible for
the slowing down in the decrease of [$\alpha$/Fe] at that same epoch.

Point (2) implies that stars in the thick disk were formed in an ISM that
was well-mixed at any given time.  It suggests that the thick disk formed
its stars in a turbulent ISM for several Gyr, confined to the inner parts
of the Galaxy \citep[compatible with a short scale length,][]{bov12b}. 
This well-mixed gas was progressively enriched, giving rise
to a monotonic and well defined age-metallicity-[$\alpha$/Fe] relation.
The result that the ISM at the epoch of the thick disk formation was well mixed is supported by the fact that no radial gradient has been detected
in the $\alpha-$enriched population, over a spatial scale of about 10 kpc
\citep[see][]{che12a}.  Moreover, if the $\alpha-$enriched population had an 
initial gradient that was subsequently  erased, we should find a significant 
increase of the metallicity dispersion with age (Minchev et al. 2012c), especially 
at ages greater than 8 Gyr, which is not observed.  It implies that even in the presence of stellar
migration, no change in the age-chemistry relation is expected, since
the enrichment has been similar throughout the thick disk.  
The small scatter in the age-chemistry relation of the thick disk also leaves 
not much room for a component resulting from a mixture of satellite-accreted and in situ stars. 
The present dataset is not adequate to sample the metal-poor end of the
age-chemistry relation, and to see how it behaves at the thick disk-halo
interface. At the metal-rich end of the correlation however (at about
8 Gyr), it can be seen that thick disk objects reach solar metallicity
and [$\alpha$/Fe]$\approx$0.1, suggesting that they provide the chemical
initial conditions of the thin disk formation. Since the ISM is well mixed
during the thick disk phase, these initial conditions may be common to
the whole inner thin disk.  Given that the radial extension of the high
[$\alpha$/Fe] population seems to have a limit at about 10~kpc from the
galactic center \citep[][]{che12a}, we consider this is also the  limit at
which the thick disk provided significant material to set up the initial
conditions from which the thin disk formed.  Note that this explains the
apparent coincidence between the step at 10 kpc in metallicity between the
inner and outer part of the thin disk and the extent in radius of the
$\alpha$-rich population. We contend that the thin disk inside 10~kpc
is in direct filiation with the thick disk, while outer parts of the thin
disk had a different and separate history.

Point (3) implies that the ``thick disk'' population is not only thick,
it has a thin component, or, more precisely, is a continuum of components
from a thick to a thinner disk, as measured from the decrease of the
vertical velocity dispersion from [$\alpha$/Fe]=0.3 to 0.12 dex.  
Note that a continuity in the [$\alpha$/Fe]-[Fe/H] plane for the inner (thin+thick) disks,
as evidenced by \citet{bov12a}, does not imply necessarily a unique star formation
regime all along the disk sequence. Bovy et al. excluded the presence of a separate
thick disk component on the basis of the continuity in the chemical properties of 
their stars. Our analysis shows that even if these properties draw a continuum, 
their related star formation history does not, as evidenced by the different slopes 
in the [$\alpha$/Fe]-age and [Fe/H]-age sequences.
Note also that an age-metallicity
correlation in the thick disk, together with an age-$\sigma_W$ relation
is expected to give a vertical gradient: older and metal weaker stars
will reach higher distances from the galactic plane.  Hence the thick
disk has a vertical \citep[see][and references therein]{kat11}, but no radial metallicity gradient.

Point (4) suggests that the metal-poor thin disk and the thick disk
stars must have
formed in two different parts of the disk, because they have similar ages and [$\alpha$/Fe] abundances,
but different metallicity and different rotation speeds.  It also suggests that,
at the onset of the thin disk formation  (at ages of about 8-10 Gyr),
stars started to form first in the outer parts (R$>$10 kpc), while the
thick disk was still forming stars in the inner parts.  At around 8 Gyr,
the formation ceases in the thick disk altogether, and after some time,
starts in the inner thin disk, following different paths of metallicity
increase, and decrease in $\alpha$ enhancement, and of course, different
intrinsic dispersion in the vertical disk direction.

We note also that the similar amount of [$\alpha$/Fe] abundance at 8-10
Gyr in the thick disk and in the outer thin disk stars suggests that the
material out of which metal-poor thin disk has formed may have been
contaminated by the thick disk nucleosynthesis, but diluted in the
outer disk gas reservoir.  Finally, the presence of these metal-poor
thin disk stars with similar characteristics in age and $\alpha$ elements
as the thick disk goes against the suggestion that these objects could
have formed also in the inner disk and then migrated to the outer disk:
being formed at similar times, they must originate from a different
lower metallicity environment.

\section{Discussion} 

In this section, we first review some critical points: the status of
the metal-poor thin disk stars, the weak evidence for significant radial
migration and inside-out formation of the disk.
We then sketch a scenario of the formation of the Milky Way disk based on our results 
and discuss how this scenario fits into a broader context.

\subsection{The status of metal-poor thin disk stars and the outer thin disk}

Different interpretations of the presence of metal-poor thin disk stars
have been envisaged, based on similar data of solar neighbourhood stars.
\citet{red03} or \citet{ben12a} proposed that the hiatus
in metallicity between metal-rich thick disk objects (at [Fe/H]$\sim$ 0.0
dex) and thin disk metal-poor stars (at [Fe/H]$\sim$-0.6 dex) was caused
by accretion of lower metallicity gas which diluted the ISM at the end of
the thick disk phase. We think this is unlikely for two reasons. First,
young thick disk stars have similar age as the oldest metal-poor thin
disk objects. Being both formed at the same time, and since the thick
disk is a population confined to the inner parts of the Galaxy, it is
most likely that these objects originated from the outer disk.  Second,
the kinematics and orbital parameters of some of these stars is highly
suggestive of an external origin: several of these objects have large
apocenter ($R_{apo} >$10 kpc), with their V-component overtaking the
LSR (V$>$0 km.s$^{-1}$), see \citet{hay08}.
In the SEGUE sample, metal-poor thin disk stars share the same properties, 
having mean galactocentric radii above 9 kpc \citep[see][Fig.~7]{bov12b}, 
and a clear positive V-component \citep[see][Fig.~3]{liu12}.

\begin{figure}
\includegraphics[width=9.5cm]{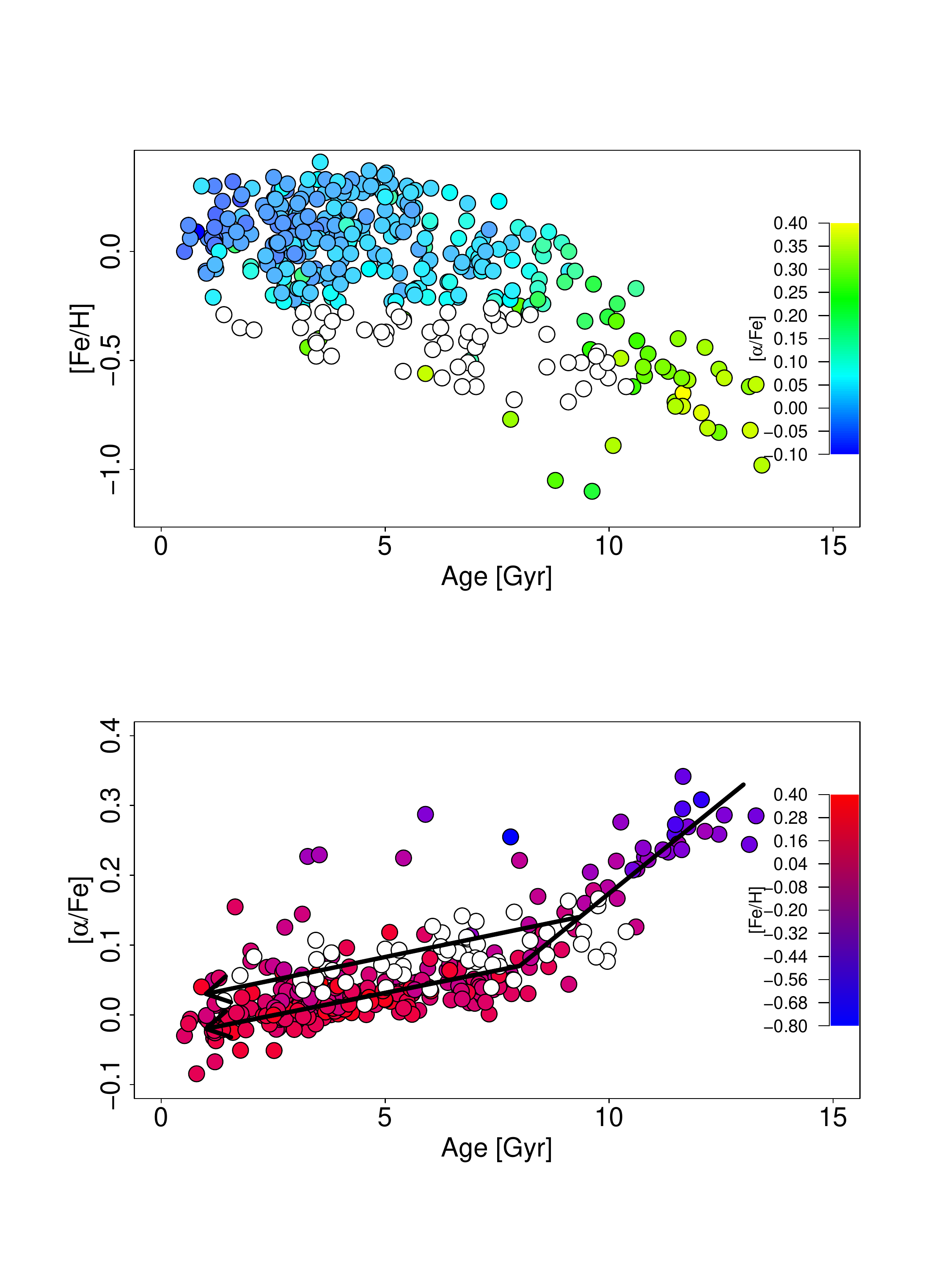}
\caption{Age-metallicity and age-[$\alpha$/Fe] distributions, with stars in the metal-poor 
tail of the thin disk sequence (Fig. 1) at [Fe/H]$<$-0.3 dex shown as white dots. 
The metal-poor thin disk stars draw a sequence parallel to the local thin disk, joining 
the thick disk at 9-10 Gyr and [$\alpha$/Fe] in [0.1,0.15].}
\label{agealpha3}
\end{figure}

These properties have to be combined with the fact that, starting at R$_{GC}$=9-10 kpc, the disk 
mean metallicity drops below [Fe/H]$\approx$$-$0.3 dex, with an 
apparent flat distribution beyond this distance. In particular, this is observed
for open clusters, which exhibit a (possibly) steeper drop in metallicity for older clusters
\citep[ages $>$ 4Gyr; see ][ Fig.~7]{bra12}, while field giants have mean metallicities of 
$-$0.48 dex and [$\alpha$/Fe]=+0.12 \citep{ben11}, in good agreement with \citet{adi12} for similar stars. 
Hence, local metal-poor thin disk stars have mean orbital radii of about 9-10 kpc and stars 
at 9-10 kpc have metallicities and alpha-element abundances similar to their local 
counterparts. Since stars of such metallicities and alpha abundances seem dominant beyond R=9 kpc, 
it suggests that metal-poor local counterparts originated in the outer disk.
Moreover, while vertical motions of field stars in the outer disk have not yet been measured, 
some open clusters are found at distances of several hundred parsecs from the galactic plane, 
which seems in line with the rather hot kinematics of local metal-poor thin disk stars (see \S~3.5.2.). 
There is therefore strong similarities of the properties of these stars with those of the outer disk as a whole.
It is this relationship, together with the fact that local metal-poor thin disk objects 
can have ages up to $\sim$10 Gyr, which suggests that the outer thin disk must have started to form stars 
before the inner thin disk.
Fig. \ref{agealpha3} highlights the position of metal-poor thin disk stars having metallicities
below [Fe/H]=$-$0.3 dex on the thin disk sequence (see Fig. \ref{alphafeh}). 
These objects form a sequence parallel to that of the local thin disk, but at higher relative $\alpha$ abundance, 
connecting to the thick disk at ages of 9-10 Gyr.

Note that a number of extragalactic studies show the presence
of old stars in the outskirts of disks \citep[][among other studies]{fer01,vla10,bar11,yoa12}. 
These stars may have been formed in the outskirts of galaxy disks, without necessarily having migrated from
the inner disk regions \citep[see, for example, the discussion in ][]{san09, gib12}.
Several studies also have shown the change in metallicity profile that occurs in the 
outskirts of disks (Bresolin et al 2012 and references therein). In the case of M33, it seems particularly clear that the increase 
in mean age beyond the break radius is accompanied by a very significant decrease
in metallicity \citep[][]{bar11}. A behavior which ressembles very much what is
observed in the Milky Way.

To some extent, we confirm the viability of [$\alpha$/Fe] as a proxy
for age used by Bovy et al. (2012b): [$\alpha$/Fe] correlates well with
age, but this correlation is stronger for the thick disk population
(Fig. \ref{agealpha2}).  However, the specific status of metal-poor
thin disk stars is not discussed in the study of \citet{bov12b}.
Metal-poor thin disk stars do not fit into their scheme because
they result from different conditions during their formation. At
[$\alpha$/Fe]$\sim$0.1 dex, they have higher vertical velocity
dispersion, hence higher scale height, than local thin disk stars with
similar relative alpha element content.  In the \citet{bov12b}
decomposition, these objects are outliers to the scale height--scale
length anti-correlation, and possibly results in the point that can
be seen at (4.3 kpc, 440pc; their Fig.~5). Furthermore, it is
tempting to suggest that it is the presence of these metal-poor thin
disk stars that leads \citet{bov12a} to conclude on the absence of
a dip between alpha-poor and alpha-rich stars: these objects have
alpha abundance intermediate between the local disk and the thick disk,
and kinematics hotter than standard thin disk. They will contribute to
fill the dip between the thin and thick disks, although they  cannot
be considered an intermediate phase between the two, and should not be
counted undistinguished when estimating surface densities. Hence, we infer
that the dip probably is real, although confirmation would require to
properly disentangle the different stellar populations in the SEGUE data.

While we discuss the conditions of the ISM in which stars may have
formed in the outer disk in \S~5.4.2, we note that such stars, having,
at similar ages, the same level of alpha abundance ratios, may be
related to the thick disk population. In a highly turbulent
disk, like the one that may have formed the thick disk, a significant
amount of gas is expected to be expelled outwards, likely contaminating
the outer regions (\S~5.4.2). Combined with metal-poor accreted gas,
this gas could have provided the mixture from which these stars formed
resulting in their observed chemical composition.  This would provide
a natural explanation of the high abundance ratios found in the outer
regions for which only adhoc solutions have been proposed until now.

\subsection{Radial migration} \label{migration}

\subsubsection{Thin disk}

While the `contamination' or `pollution' of the solar vicinity by stars
originating from other radii hampers direct deciphering of the evolution
of our Galaxy from simple correlations between ages and chemical properties
of ensembles of stars, such contaminants do convey crucial information about the formation of the
disk on larger scales.  Since the suggestion by \citet{sel02}
that migration could play a significant role in redistributing
stars in disks, there has been several studies suggesting that migration
is necessary to explain patterns observed in the solar neighbourhood
\citep[][]{hay08,sch09a,sch09b,loe11}.

We distinguish two cases: 

\begin{itemize}

\item Limited migration, in which case we consider that only the tails of the solar
neighborhood metallicity distribution (at [Fe/H]$<$-0.2 dex and [Fe/H]$>$+0.2 dex) 
are populated by stars that come from other radii. In this case, which mechanism 
(churning/blurring) is more likely to have produced such contamination? 
 These contaminating stars have metallicities similar
to stars found at 2-3 kpc from the Sun, and which have a mean metallicity
of +0.2 dex at just 2~kpc towards the galactic center \citep{hil12},
and -0.3~dex at 10~kpc \citep{ben11}.  Simple estimates of radial excursions
due to epicycle oscillations give of the order of 1-2~kpc at the solar
galactocentric radii, which put stars of the inner or outer disk within
reach of the solar vicinity. These stars are found in limited amounts in
the solar neighborhood (a few percent), and certainly does not require
significant radial migration across the solar circle. If only the
tails require pollution by objects born outside the solar circle, then
probably epicycle oscillation could bring a few percent of stars that
dominate the disk at just 2 kpc from the solar radius. 
We consider that blurring (which we know to occur in disks) would be sufficient to explain most of these tails, 
even though we do not exclude that a limited amount of churning may have contributed.
Orbital parameters of metal-poor thin disk stars support this conclusion.

\item Strong migration, where it has been suggested that the mean
metallicity of solar neighborhood stars is dominated by objects that
have migrated from the inner disk \citep[see, e.g ][]{loe11}. This argument has been proposed in
particular to explain how the mean metallicity at the solar radius had
already reached [Fe/H]$\sim$$-$0.1 dex 8-10 Gyr ago. However, our results show 
that the thick disk stars set the chemical initial conditions for the formation 
of the thin disk, hence there is no need to invoke the action of radial migration 
to explain the metallicity of the old thin disk (i.e., by bringing stars with 
[Fe/H]$\sim$$-$0.1 dex into the solar neighborhood). 
Moreover, the fact that the outer disk is dominated by stars with a
metallicity significantly lower than that of the youngest thick disk
means that no inner disk stars (either thin or thick disk) have
migrated to the outer disk in significant numbers over the last 10 Gyr.

\end{itemize}

Therefore, our conclusion is that essentially no radial migration (in the sense of
churning) is necessary to explain the characteristics of the sample
of stars studied here.  Rather, it is the position of the Sun at
the interface of the inner and outer disk that results in the local
distribution of stellar abundances having significant tails at high and
low metallicities.

Finally, it is interesting to speculate about the radial age-profile
which arises from the picture outlined in \S~5.1 and from the results
discussed in \S~4.  In the inner disk (R$<$ 10~kpc), the superposition of
an old thick disk with a short scale length \citep[2kpc; ][]{bov12b}
and a thinner and younger component with a long scale length (3.6 kpc)
will naturally give rise to a decreasing mean age outward.  In the
outer regions (R$>$ 10~kpc) where stars older than local thin disk
stars can be found, the mean age could be higher than that within
the solar neighborhood, giving rise to a U-shape age profile, without
invoking any significant mixing \citep[]{ros08}. Of course,
this conclusion depends on the way star formation proceeded in the
outer disk of the Milky Way, but at least suggests the possibility
that U-shaped age profile could exist in disks where there has been no
substantial redistribution of stars due to migration.  This reinforces
the fact that, even in $\Lambda$CDM scenarios for galaxy formation and
evolution, old stars may have been formed in the outskirts of galaxy
disks, without necessarily having migrated from the inner disk regions
\citep[see also the discussion in ][]{san09}. 
Understanding where these stars have formed is crucial in
shedding light on the early epochs of thin disk formation.

\subsubsection{Thick disk}

It is also interesting to discuss the impact that radial migration may
have had in forming the Milky Way thick disk population, in light of our age
estimates of thick disk stars. The possibility that the thick disk is the result of secular processes
instead of more violent ones has been extensively debated
in the last couple of years \citep[][]{sch09b, loe11, min12c}.
The first consideration to make is that
radial migration, as it has been discussed in the literature  
\citep[]{sel02, sch09a, ros08, min10, min11, min12b, min12c, dim13} is the result
of a secular process related to the presence of stellar
asymmetries likes bars and spiral arms in galaxy disks. Similar structures in disks are
almost absent from the progenitors of present-day spirals at z $>$1.5
and they generally remain rare up to $z\sim1$ \citep[e.g., ][]{she08}. This epoch is perhaps 
the transition between an early violent phase of disk formation to
a more quiescent one \cite[e.g.,][]{mar12, kra12}. The
redshift evolution of the bar fraction
seems to agree with this scenario, indicating that the bar fraction drops
by a factor of about three from z = 0 to z = 0.8  \cite[][]{she08}.  If stellar asymmetries
were rare above $z\sim1$, redshifts higher than this do not correspond
to a secular evolution phase for galaxy disks, we have to conclude that if radial
migration formed thick disks, this can have been possible only in the
last 8 Gyr of galaxy evolution. Within this context, it is perhaps surprising
then 
to see that thick disk stars in the solar neighborhood are \emph{all}
older than 8 Gyr (as indicated for example by the $z_{max}-$age and $W-$age relations). If radial
migration only started to affect galaxy disks during the last 8 Gyr,
producing the observed thickening of the disk at the solar vicinity,
we should expect to find a significant fraction of stars with ages
less than 8 Gyr with high $z_{max}$ and W. There is no trace of these
stars within the solar vicinity. All stars with ages younger than 8 Gyr
are redistributed in a thin configuration ($z_{max} < $0.5 kpc). This,
in our opinion, rules out the possibility that migration may have been a
major contributor to the thickening of the galactic disk. 
Moreover, during the evolution of the thick disk there was a
decrease in the disk scale height which is not accompanied by an
increase in the disk scale length \citep[][]{bov12b}. This favors scenarios which cool
the disk vertically without substantially heating it
radially. Once again, radial migration seems incompatible with these
findings -- migration is usually accompanied by an increase in
the disk scale lengths, with a net outward flow of stars (i.e. radial heating).

Apart from the thickening of a pre-existing thin disk component,
cosmological numerical simulations have often claimed that radial
migration is a fundamental mechanism in driving the evolution  of a thick
disk  because, in  these models, thick  disk stars within the solar
neighborhood are found with a variety of formation radii, frequently
smaller than 4 kpc \citep[][]{bro12, min12c}. 
A  small formation radius is thus associated with subsequent radial
migration which would carry those inner thick disk stars out to the
solar annulus. Note however that if thick disk stars formed already in a
thick disk configuration with high velocity dispersions and an angular
momentum content lower  than that of the thin disk (Figs.~12, 13,
14), they must have had eccentric orbits \emph{already} at the time of
their formation.  An eccentricity  $e=0.3$ would be sufficient to allow
a star born at R=4kpc, for example, to reach the solar annulus, without
any need to invoke strong subsequent migration.  In \citet{bro12}
for example, most thick disk objects have formation radii in the inner
disk (see their Fig.~9), but moderate eccentricities would be sufficient
to bring a substantial number of these objects out to the solar radius.
The recent models by \citet{bir13} seem to support this scenario:
their thick disk  stars have small formation radii, and their  guiding
centers  do not  change significantly  over  time, thus indicating
that thick disk stars had already eccentric orbits at the time of their
formation. In \citet{bro12} also, the thick disk forms with a scale
length of 1.7 kpc, very near to the observed value of 1.8 kpc \citep[][]{bov12b}. Thus one may wonder if any substantial migration is needed
to explain the observed properties of the thick disk in the Milky Way.

\subsection{Inside-out formation of the disk}

An interesting consequence of the proposed scenario concerns the so-called
inside-out paradigm that has been proposed for the formation of disks
\citep[Larson 1976, and e.g.][]{som03,rah11,bro12}, and which received support from the analysis of \citet{bov12b}.  
How much of that picture is due to the superposition
of two components that have different scale lengths and ages, and how much
to the effect of a real inside-out process?

Combining our results with those of \citet{bov12b}, one
can see that the formation of the thick disk, which lasted 4-5 Gyr
(Fig.~\ref{agealpha2}), did not produce any significant increase in scale
length -- the scale length remains $<$ 2kpc down to [$\alpha$/Fe]$\sim$
0.25, which, in their $\alpha$ abundance scale, is the beginning of the thin
disk regime (see their Fig. 5). At the same time, the thick disk scale
height decreased by a factor of 2 to 3.  Moreover, a tight age-metallicity
relation as found here, combined with an inside-out scenario, should
produce a radial metallicity, or [$\alpha$/Fe], gradient in the thick
disk, which is not observed \citep[][]{che12b}.

Within the (inner) thin disk itself, whose formation lasted 8 Gyr, one
can hardly see any trend in support of an inside-out formation (blue and cyan
points in the upper panel of Fig. 5, or orange and red points in the
lower panel, \citet{bov12b}): several points with the lowest [$\alpha$/Fe] ($<$0.05 dex),
and high metallicities, which ought to be young and to have long scale
length in the scheme of Bovy et al., have indeed scale lengths as short
as 2.1 or 2.8 kpc, while some others, at the limit between the thick and
thin disk in terms of [$\alpha$/Fe], have scale lengths greater than 4 kpc.

It is clearly the combination of the two structures that gives Bovy et
al. the impetus to favor an inside-out formation scenario, but this is much less
supported once one looks more carefully into the details of the age structure of
the population.

Finally, it is worth commenting on some of the implications that
result from assuming that the formation of the disk proceeded inside-out.
\citet{bov12b} proposed that the outer disk formed after the
inner disk because of it's larger scale length and lower metallicity, which 
were interpreted as a signature of younger ages \citep[from the model of ][]{sch09b}.
Similarly, \citet{ros13} predict that the outer disk should be essentially younger
than 2 Gyr following  \citet{bov12b}.  Clearly, the observations are
in conflict with these predictions: open clusters older than 4-5~Gyr are
seen beyond 11~kpc (Berkeley 17, 32, 36, 39) and out to at least 15~kpc
(Berkeley 20, 29), while metal-poor thin disk stars seen in local samples
are older than the general population of local thin disk stars. 
All these arguments suggest a more complex star formation history than that predicted until now - simple inside-out 
formation scenarios and radial-mixing alone do not alleviate this tension between models and observations.

\subsection{Proposed scenario and Context}

\subsubsection{Qualitative outline of the evolution of the MW}

From the results presented in \S~\ref{res} and \ref{imp} we outline
the following qualitative scenario: The first stars in the thick disk
started to form out of a turbulent gaseous disk at redshift z$>$3.
This gaseous disk was characterized by high velocity dispersions, had a
short scale length \citep[][]{bov12b} and large scale height ( deduced
from the $z_{max}$ \emph{vs} age and $\sigma_W$ \emph{vs} age relations;
Fig.~\ref{wsigalpha}), and a low rotational velocity. The strong feedback
associated with an active phase of intense star formation at that epoch
was likely responsible of an efficient mixing of metals in the ISM as
suggested by the tight age-metallicity relation found in local thick disk
(\S~3.3), and in the absence of a radial gradient among thick disk stars
over several kpc \citep[][]{che12b}.  We suggest that the increase
in the stellar velocity dispersion that the data show for ages greater
than $\sim$8 Gyr is related to the fact that, at those times, the Milky
Way disk had high star-formation intensities, similar to local starburst
galaxies, i.e. with star formation rate densities $\Sigma_{\rm SFR} >$
0.01 M$_{\odot}$ kpc$^{-2}$ yr$^{-1}$ \citep[e.g. ][]{leh96, meu96}.

We concluded the star formation in the early Milky Way must have been
intense through the following considerations.  \citet{bov12b} found a thick disk
scale length of the order of 1.8 kpc. Assuming that the thick disk has
an exponential density distribution, and a local volume density
ratio between thick and thin disk stars of 0.05 (which is at the lower end
of possible values, see \citet{cha11}, then, the thick disk mass
is no less than two tenths of the thin disk mass. We assumed the scale height
ratio between the two populations is 1/4 and that both disks have similar scale lengths,
but the thick disk mass could be much higher (perhaps up to the thin disk mass $\sim$
$10^{11}M_{\odot}$), depending on the exact scale length of the thin disk.
Thick disks with comparable masses to that of their thin disks have been
found recently in some external galaxies \citep{com11}.
In case of a thick disk with just two tenths of thin disk mass (which
we believe to be a very conservative choice) and which formed over
$\sim$4 Gyr leads to an average star formation rate $<SFR>\sim$5
M$_{\odot}$ yr$^{-1}$ or an average star-formation rate per unit
area\footnote{$\Sigma_{SFR}=<SFR>/Area$. Following \citet{dib06}, we
have evaluated the galactic surface area as Area=$\pi(3r_d)^2$, with $r_d$
 being the thick disk scale length, i.e. $r_d$=1.8 kpc \citep{bov12b}} 
$\Sigma_{SFR}\sim$0.05 M$_{\odot}$ yr$^{-1}$ kpc$^{-2}$.
Such a value for the star formation intensity, which is likely a lower limit, would
imply that Milky Way would have a star formation intensity similar to
that of nearby starbursts and distant intensely star-forming galaxies
at z$\sim$2-3 \citep[e.g. ][]{leh96, dib06, leh09, leh13}
Above these values, there is a sharp increase in the
velocity dispersion of the ISM with $\Sigma_{SFR}$. We thus suggest that
the relations found between stellar vertical velocity (or $z_{max}$) and
age reflects similar relations in the ISM: stars formed in a thick ISM,
whose high dispersions were sustained for several Gyr by intense star
formation (but notice not necessarily high star formation rates overall).

Several processes may have contributed creating an initial gas rich phase
and/or in maintaining such high SFRs over several Gyr in the Milky Way
disk: multiple accretion of gas-rich companions \citep[][]{bro04, hou11}, gas 
accretion from filaments \citep[][]{ker09}, gravitational instabilities \citep[][]{for12},
gas cooling in the halo, and recycling of the gas from outflows. Heating
during the merger epoch \citep[][]{qui93, qu11, hou11}, 
or scattering of stars by massive clumps \citep[][]{bou09}
may have caused additional thickening of the already formed thick stellar
component \citep[][]{qui93, qu11}.

The star formation must have continued for 4-5 Gyr with each
new generation of stars forming in thinner and thinner layers, and increasingly supported by rotation.  During
this phase turbulent mixing of the gas maintains a homogeneous ISM,
and permits a monotonic chemical enrichment as the stellar population
evolves and enriches the ISM with metals.  The steep gradient in the
Fe enrichment with time indicates that star formation proceeded at a
sustained rate in the inner parts, but the star formation in the outer
parts must have been less intense and reached more moderate metallicities
and $\alpha$-enrichment compared to the inner parts.  At the end of
the thick disk phase ($\sim$ 8 Gyr), the inner disk is left with a
metallicity $\sim$$-$0.1$\pm$0.1~dex.  There may be an age gap in the
star formation possibly causing the gap in $\alpha$ element enrichment
\citep[][]{adi12}, after which, star formation proceeds in
a thin disk.  The forming thin disk inherited
its metallicity and $\alpha$ element enrichment level from (approximately)
the end of the formation of the thick disk.  The growth of the thin disk
must have had characteristics somewhat different from that of the thick
disk as there is a change in the chemical enrichment rate in the inner
disk.  During these periods of formation, the star formation continues
in the outer disk, as revealed by the presence of metal-poor stars
($-$0.7$<$[Fe/H]$<$$-$0.3) at all ages between 2-10 Gyr in the solar vicinity.
Being at the edge of the inner disk, the solar vicinity is dominated by
objects formed along the evolutionary path followed by the inner disk,
and is only weakly contaminated by objects that formed in the outer disk,
and which have metallicities [Fe/H]$<$$-$0.3 dex.

\subsubsection{Scenario within the context of distant galaxies}

As explained in the previous section, if the thick disk was formed in
the Galaxy, as our results suggest, most probably the Milky Way was in
starburst mode at z=2-3.  Actively star forming galaxies at high redshift
are known to be both gas rich, have high gas mass surface densities \citep[][]{dad10, ara10, dan09, tac10, comb11},
have high velocity dispersions in their
gas and intense star formation \citep[][]{leh09, leh13, swi11}. Within the context of gas rich distant intensely star forming
galaxies, global disk instabilities may have played an important role in
regulating star formation in the early universe.

In this spirit, the Toomre criteria may provide a simple way to understand
the early evolution of disk of galaxies and their global star formation
properties.  The effective Toomre criteria is given by $\frac{1}{{\rm
Q}_{\rm total}}$=$\frac{1}{{\rm Q}_{\rm stars}}$ + $\frac{1}{{\rm Q}_{\rm
gas}}$ \citep[][]{wan94, jog96, rom11}, where
for the gas, Q$_{\rm gas}$=$\kappa\sigma_{\rm gas}$/$\pi$G$\Sigma_{\rm
gas}$ and similarly for the stars, where $\kappa$ is the epicyclic
frequency, $\Sigma_{\rm gas}$ is the gas mass surface density, $G$
is the gravitational constant, and $\sigma_{\rm gas}$ is the velocity
dispersion of the gas.  The line of stability against star-formation is
Q$_{\rm total}\sim$1 where values higher than about 1 imply that a disk
is stable against star formation.  In the early evolution of galaxies,
the surface mass densities of the disks may be dominated by the gas and
are much higher than local galaxies \citep[cf.][and references therein]{dad10, tac10, ken12}. This suggests
that to maintain galaxies at about Q$\sim$1 requires higher dispersions
than local galaxies.  The thickness of the disk is proportional to
$\sigma^2$/$\Sigma_{\rm total}$.  With the higher dispersions necessary
to keep the disk near the line of stability suggests that the disks are
generally thicker than disks in the local universe as perhaps observed
\citep[e.g. ][]{elm09, epi12}.

To sustain the galaxies for the necessary length of time as observed
in the thick disk of the MW, approximately a few Gyr, some process or
processes must regulate the star formation to either equilibrate with
the gas supply, say from cosmological gas accretion or mergers, or have
star-formation self-regulated through its own energy output or combinations
of all of the above \citep[e.g.][]{elm10}.  Processes which may
drive the large scale turbulence in the ISM whether it be gravitational
collapse and instabilities on large scales, shearing from large scale
motions, energy injection from massive stars, or cosmological gas
accretion, the velocity of the cold molecular gas out of which stars form,
will tend to moderate such that galaxies are near the line of instability
in gas rich disks.  If the dispersions are too low and Q$<<$1, the star
formation intensities will grow rapidly, consuming all the gas in few
rotational or crossing times or the young stars will inject sufficient
energy to increase the dispersions in the gas tending to stabilize the
disk. If Q$>>$1, star formation will be strongly suppressed globally,
allowing the gas content to grow until the disk again becomes unstable.
The homogeneity in the stars and the smooth growth in the metallicity
and decline in the $\alpha$-enhancement suggests that thick disk must
have been mostly near the line of instability throughout its evolution,
maintaining a high level of turbulence, and supporting a thick disk whose
thickness decreased with time.

The high dispersions necessary to maintain this equilibrium in a
thick disk also has the advantage of leading to a short mixing time.
The turnover time of large scale turbulence is H/$\sigma_{\rm gas}$ which
in the case of a 1 kpc thick disk and a dispersion of 100 km s$^{-1}$
\citep[which are rough characteristics of high redshift star forming galaxies;][]{leh09, leh13}, 
leads to a timescale of 10 Myrs
which is a about a tenth of a rotation times, for a rotation speed of 150 km s$^{-1}$
and at r$_{\rm e}$=2.5 kpc \citep[about the rotation speed and half light radius of the thick disk;][]{for06,for09} 
and similar to the evolutionary timescale of
massive stars that enrich the ISM with metals through stellar winds and
supernovae. Thus the metals lost to the ISM by massive stars will mix
very quickly with some being lost to outflows because of the intense
star formation \citep[e.g.][]{kor12}.

As the gas mass surface densities drops due to the gas being consummed
by star formation and as the mass surface density of the thick stellar disk
increases, we would expect the gas disk to grow thinner -- at lower gas
mass surface densities, the velocity dispersion of the gas necessary
to maintain the gas along the line of stability drops and naturally
leads to a shorter scale height (which is proportional to $\sigma^2$).
In other words, the increasing (both in absolute terms and in terms of
declining gas fraction) stellar mass tends to stabilize the disk against
intense star formation. This stabilization is what leads to a declining
scale height with time and likely leads to greater inhomogeneities in
the abundances of the stars forming in progressively lower scale heights.

In addition, such a scenario might explain why the outer disk appears
to have characteristics of the younger stars in the thick disk.  In the
outer disk both the gas mass surface densities are lower, the gas surface
fractions ($\Sigma_{\rm gas}$/$\Sigma_{\rm total}$) are likely higher,
and the orbital time much longer.  The star formation rate per unit
area is probably also low in the outer disks 
\citep[as is observed both in local and distant galaxies, e.g.][]{ken12, leh13}.
Thus both the enrichment and the mixing timescales will be
much longer since the star formation occurs at a much lower intensity.
This will naturally lead to less energy injection from massive stars,
the accretion rate per unit area is likely lower, and the impact on a
gas rich outer portion of the disk may be less (mostly driving the gas
inwards) than in the inner disk where the gas is stabilized by the higher
stellar mass surface densities. In addition, the intense star formation
in the inner region during the early evolution of the thick disk will
disperse metal-rich, alpha-enhanced gas into the halo which then can
subsequently cool and rain back down on the disk. This would account for
the enhanced alpha-element ratios of the stars in the outer disk despite their
relatively inefficient star formation.

\section{Summary}

We summarize our results and conclusions as follows: 

(1) The thick disk arose from a well mixed turbulent gaseous disk which
gave rise to a steep and monotonic chemical enrichment lasting 4-5 Gyr.

(2) The vertical velocity dispersion progressively decreased in the
thick disk, from about 50km s$^{-1}$ to 25km s$^{-1}$, suggesting that
star formation proceeded in progressively thinner layers.

(3) The {\it inner} thin disk ($<$10 kpc) inherited from the chemical
conditions left at the end of the thick disk phase, some 8 Gyr ago.
The transition between the two disks is imprinted in the fossil signature
left by the change of regime that can be seen in the rate of [$\alpha$/Fe]
variation as a function of time (Fig. \ref{agealpha2}). This is the only
discontinuity that we can detect between the two populations.

(4) Combined with the scale length and scale height measurements from \citet[][]{bov12b}, 
these results do not support a continuous inside-out
formation of the disk.  We advocate that it is only the combination of
two structures with different scale lengths that gives the impression of
an inside-out process, but that there is hardly any trace of such
process in each population, although their formation lasted several Gyr.

(5) The outer thin disk started to form stars while the thick disk phase
was still on-going in the inner disk, 9-10 Gyr ago, possibly from a 
mixture of enriched material expelled from the thick disk, and accreted gas depleted of metals. 
However, the evolution of the outer disk remained essentially disconnected from that of the inner disk afterwards.

(6) Combination of points (1-4) leaves little room for radial migration, in the sense
of churning, to have played a substantial role in redistributing stars seen in the 
solar vicinity, or across the solar annulus.

Other factual results have been obtained :

(7) The rate iron enrichment in the thick disk was about 5 times higher as compared to
the thin disk phase.

(8) The vertical velocity dispersion of old disk stars is a mixture of objects of different 
origin: the local thin disk ($\sim$ 8 Gyr) and the young thick disk (9-10 Gyr), which, although being of 
different ages, have the same vertical velocity dispersion (around 22-27 km.s$^{-1}$), probably due
to being heated by the same process, and the old metal-poor thin disk (9-10 Gyr), which although having 
the same age as the young thick disk, have higher vertical velocity dispersion ($\sim$ 35km.s$^{-1}$).
It is the mixture of different amount of each of these populations in studied samples 
which have conducted to varying results.

Finally, we conclude with a point of nomenclature.
The search for a parameter most closely related to the defining properties and limits of a stellar population 
has questioned galactic astronomy for at least fifty years \citep[e.g.][]{sch58}. With this work we went one step further in this direction by linking the gross 
characteristics of a population to properties that reflect the ISM in which 
it was formed at an identifiable epoch.
In doing so however, we pulled the old labelling of ``thin'' and ``thick'' disks further 
away from their original meaning. 
In our new definition, the thick disk is not only thick, 
it has a thin component, it is not only older than, but also coeval with the (outer) thin disk. The ``thin'' 
outer disk, as sampled in the solar vicinity, shows kinematics more akin to a conventional ``thick'' disk. 
While we have maintained throughout the paper
the conventional designation of thick and thin disks, other designations could be more appropriate. 
There is perhaps more continuity between the thin and thick disks than between the inner thin disk and outer 
thin disk. The inner thin disk and the thick disk seem to be more  like the same structure and could be 
designated as the young and old inner disk, while the outer disk appear more like a separate component.

\begin{acknowledgements}
MH wish to thank Vardan Adibekyan for his kind advice in the use of the data, and Poul Nissen for interesting comments on the
first version of this article. 
The authors acknowledge the support of the French Agence Nationale de la Recherche (ANR)  under contract ANR-10-BLAN-0508 (Galhis project)
and wish to thank the referee for helpful and constructive comments which greatly improved this paper.
This research has made use of the SIMBAD database, operated at CDS, Strasbourg, France 
\end{acknowledgements}

\end{document}